\documentclass[pre,superscriptaddress,twocolumn]{revtex4-1}
\pdfoutput=1
\usepackage[colorlinks=true,urlcolor=blue]{hyperref}
\usepackage{amsfonts}
\usepackage{graphicx}
\usepackage{mathrsfs}
\usepackage{amsmath}
\usepackage{amssymb}
\usepackage{xcolor}
\usepackage{bbold}
\usepackage{bm}

\usepackage{color}
\definecolor{red}{rgb}{0.75,0,0}
\definecolor{blue}{rgb}{0,0,0.75}
\definecolor{green}{rgb}{0,0.5,0}

\begin{document}

\title{Nonreciprocity as a generic route to traveling states}
\author{Zhihong You}
\email[Corresponding author:\ ]{youz@ucsb.edu}
\affiliation{Department of Physics, University of California Santa Barbara, Santa Barbara, CA 93106, USA}
\author{Aparna Baskaran}
\affiliation{Martin Fisher school of Physics, Brandeis University, Waltham, MA 02453, USA}
\author{M. Cristina Marchetti}
\email[Corresponding author:\ ]{cmarchetti@ucsb.edu}
\affiliation{Department of Physics, University of California Santa Barbara, Santa Barbara, CA 93106, USA}

\begin{abstract}
We examine a non-reciprocally coupled dynamical model of a mixture of two diffusing species. We demonstrate that nonreciprocity, which is encoded in the model via antagonistic cross diffusivities, provides a generic mechanism for the emergence of traveling patterns in purely diffusive systems with conservative dynamics. In the absence of non-reciprocity, the binary fluid mixture undergoes a phase transition from a homogeneous mixed state to a demixed state with spatially separated regions rich in one of the two components. Above a critical value of the parameter tuning non-reciprocity, the static demixed pattern acquires a finite velocity, resulting in a state that breaks both spatial and time translational symmetry, as well as the reflection parity of the static pattern. We elucidate the generic nature  of the transition  to traveling patterns using a minimal model that can be studied analytically. Our work has direct relevance to nonequilibrium assembly in mixtures of chemically interacting colloids that are known to exhibit non-reciprocal effective interactions, as well as to mixtures of active and passive agents where traveling states of the type predicted here have been observed in simulations. It also provides insight on transitions to traveling and oscillatory states seen in a broad range of nonreciprocal systems with non-conservative dynamics, from reaction-diffusion and prey-predators models to multispecies mixtures of microorganisms with antagonistic interactions. 
\end{abstract}

\maketitle

Traveling patterns occur ubiquitously in nature. Examples range from oscillating chemical reactions~\cite{Field1985,Weiss2017,Semenov2016}, waves of metabolic synchronization in yeast~\cite{Schuetze2011}, to the spatial spread of epidemics~\cite{Diekmann1978,Diekmann1979,Abramson2003,Wang2013}. Most mathematical models that capture such spatio-temporal dynamics, including reaction-diffusion equations~\cite{Field1985,Erneux1993,Krischer1994,Kondo2010,Gambino2013}, excitable systems~\cite{Keener1980, Holden2013}, collections of coupled oscillators~\cite{Hong2011, Hong2012}, and prey-predator equations~\cite{Tsyganov2003,Biktashev2009,Mobilia2007} are unified by the fact that the dynamical variables are \emph{non-conserved} fields~\cite{Coullet1989}. In this case  the coupling to birth-death or to other reaction processes provides a promoter-inhibitor mechanism that sets up oscillatory states. In this paper we demonstrate that  traveling patterns can arise in  multi-component systems described by purely diffusive \emph{conserved} fields from non-reciprocal interactions between species. The appearance of traveling or sustained oscillatory states in a purely diffusive system with no apparent external forcing is unexpected and defies intuition.  Our work suggests that non-reciprocity provides a generic mechanism for the establishment of traveling states in the dynamics of conserved scalar fields.

The third law of Newtonian mechanics establishes that interactions are reciprocal: for every action there is an equal and opposite reaction. While of course this remains true at the microscopic level, non-reciprocal \emph{effective} interactions can occur ubiquitously on mesoscopic scales when interactions are mediated by a nonequilibrium environment~\cite{Ivlev2015,Durve2018,Hayashi_2006,Paoluzzi2018, Paoluzzi2020}. A striking physical example is realized in diffusiophoretic colloidal mixtures~\cite{Soto2014,Agudo-Canalejo2019,Saha2019}. Non-reciprocal interactions are also the norm in the living world. Examples are promoter-inhibitor interactions among different cell types~\cite{Theveneau2013} and the antagonistic interactions among species in bacterial suspensions~\cite{Pigolotti2014,Long2001,Weigel2020,Yanni2019,Curatolo2019}. Social forces that control the behavior of human crowds~\cite{Helbing1995,Helbing2000,Bain2019} and collective animal behavior~\cite{Strandburg-Peshkin2013,Vicsek2012} are other important examples as well.

\begin{figure}[t]
\centering
\includegraphics[width=\linewidth]{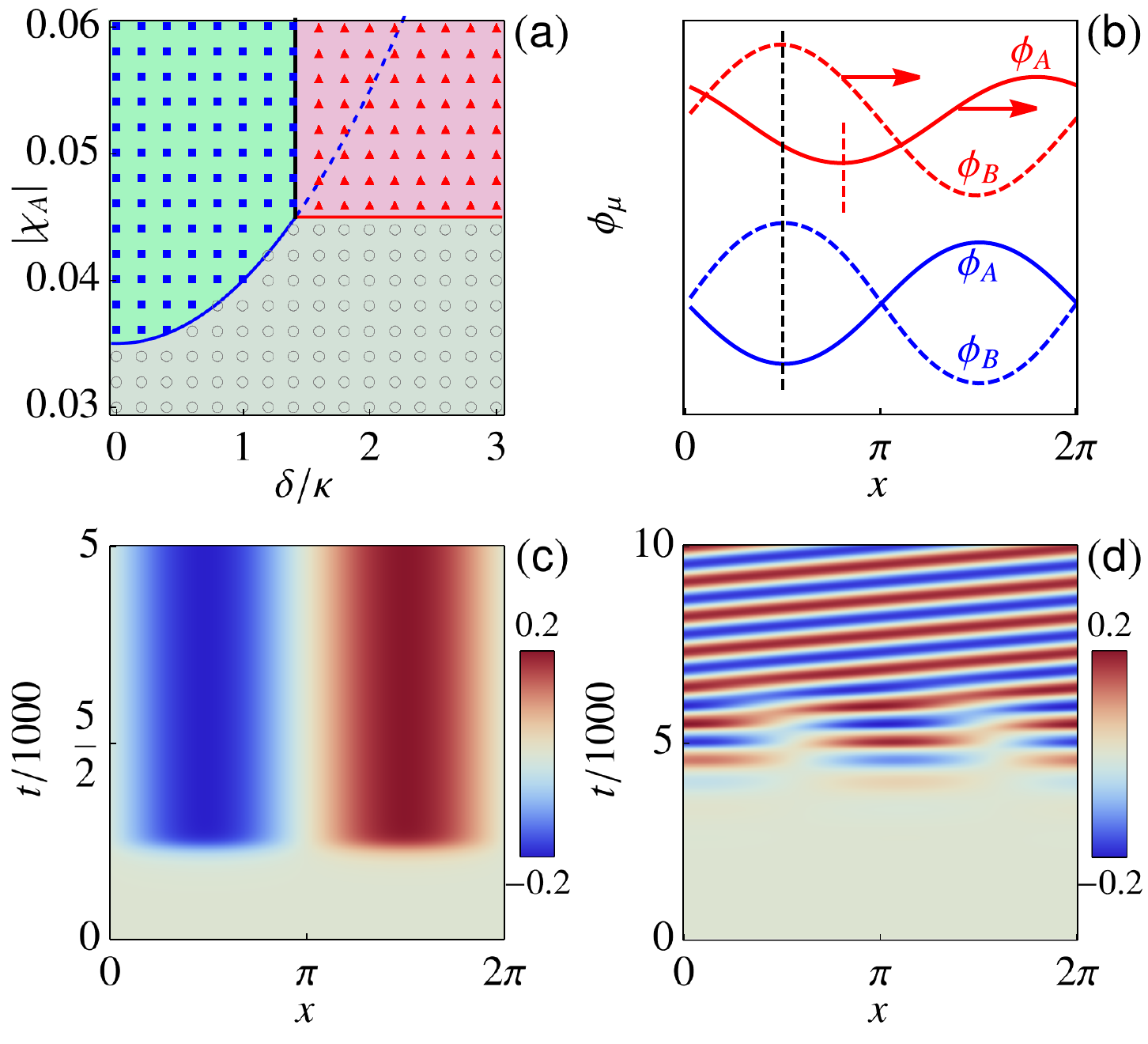}
\caption{\label{fig:Fig1} (a) State diagram spanned by $\delta/\kappa$ and $\chi_{A}$. The system has three distinct states: homogeneous (gray, circles), static patterns or demixed  (cyan, rectangles), and traveling patterns (pink, triangles). Symbols indicate results from the simulations, while the lines marking the boundaries of the colored domains are obtained from the stability analysis of the one-mode model. The static demixed state exists in the region between the black and the dashed blue lines, but is unstable. (b) Examples of spatial variations of $\phi_{A}(x)$ (solid lines) and $\phi_{B}(x)$ (dashed lines) in the static (blue) and traveling states (red). (c--d) Spatiotemporal patterns of $\phi_{A}(x,t)$ in the (c) static and (d) traveling states. In (b--d), we use $\chi_{A}=-0.05$ and (c) $\delta=\kappa$ and (d) $\delta=2\kappa$. }
\end{figure}

To highlight the role of non-reciprocal couplings in driving time-dependent phases, we examine a minimal model of the dynamics of  two interdiffusing species, each described by a scalar  field $\phi_\mu$, for $\mu=A,B$. The evolution of each concentration field is governed by a $\phi^4$ field theory that allows for a spinodal instability according to Model B dynamics ~\cite{Hohenberg1977}. When decoupled, each phase field can undergo a Hopf bifurcation describing the transition from a homogeneous state to a phase-separated state composed of  dilute and dense phases. The two fields are coupled via cross-diffusion terms with diffusivities $\kappa_{\mu\nu}$. When these couplings are reciprocal,  the interaction between the two fields  leads to a transition between a  mixed state where both fields are homogeneous to a demixed state with distinct regions of  high $A$ and low $B$. Non-reciprocity is introduced by allowing the two cross-diffusivities to have opposite signs and is quantified by $\delta=(\kappa_{BA}-\kappa_{AB})/2$.  Non-reciprocal cross-diffusivities drive a second transition through a drift bifurcation to a time-dependent state that breaks parity, where the domains of the demixed regions travel at a constant drift velocity.  This transition is closely related to ones previously reported in specific models of prey-predator and reaction-diffusion dynamics~\cite{Fauve1991,Erneux1993,Krischer1994,Tsyganov2003,Biktashev2009,Kondo2010,Wang2013}, but occurs here from the coupling of two \emph{conserved} fields. We demonstrate that the transition to traveling states is a parity and time-reversal (PT) symmetry breaking bifurcation that arises \emph{generically} from non-reciprocal couplings. The phase diagram obtained from numerical solutions of a one-dimensional realization of this minimal model in the simplest case where only  field A is supercritical, while B is subcritical, i.e., the ground state value of field B is simply $\phi_B^{g}=0$, is shown in Fig.~\ref{fig:Fig1}a.   Tuning the control parameter that drives phase separation of species $A$ ($\chi_A$) and the measure of non-reciprocity $\delta$, we observe three distinct states: a mixed state where both fields are homogeneous, a static demixed state that breaks translational symmetry with out-of-phase spatial modulations of the two fields, and a time-dependent state that additionally breaks reflection and time-reversal symmetry, where the spatial modulation of the demixed state travels at constant velocity. The solid lines are obtained from a one-mode approximation to the continuum model that can be solved analytically and provides an excellent fit to the numerics.  Within this one-mode approximation,
the transition from the stationary  to the traveling state can be understood as an instability of the relative phase of the first Fourier harmonic of the fields. The instability arises because non-reciprocity allows perturbations in the two fields to travel in the same direction, promoting a ``run-and-catch'' scenario that stabilizes the traveling pattern. While the spatial pattern in the static demixed phase is even in the relative displacement of the two phase fields, non-reciprocity breaks this reflection symmetry in the traveling state, mediating a PT-symmetry-breaking transition. Note that the transition to a PT-broken phase occurs at finite value of $\delta$, hence requires sufficiently strong non-reciprocity.  Finally,
the phase boundary separating the static and traveling patterns in Fig.~\ref{fig:Fig1}a corresponds to a so-called ``exceptional point'' where the eigenmodes of the matrix controlling the dynamical stability of the system coalesce~\cite{Kato1966,Bender2007,Fruchart2020}.
In parallel to our investigation, Saha et al.~\cite{Saha2020} have  also reported traveling density waves in  scalar fields with Cahn-Hilliard dynamics and non-reciprocal couplings. The simulations carried out by these authors support our finding that  non-reciprocal couplings provide a generic mechanism for breaking time-reversal symmetry and setting spatial patterns in motion.

A microscopic model that displays the phenomenology captured by Fig.~\ref{fig:Fig1}a is a mixture of active and passive Brownian particles, where the active component exhibits motility-induced phase separation and fluctuations in the density of passive particles  can enhance fluctuations in the density of the active fraction via an effective negative cross diffusivity~\cite{Wysocki2016,Wittkowski2017,Agrawal2017}.  The connection between the active-passive mixture and the dynamics embodied by our model is  unfolded in the SI. Another realization of this macrodynamics is a binary suspension of colloidal particles where species A attracts species B, but species B repels species A. Such competing interactions have been studied in simple models~\cite{Yllanes2017,Bonilla2019} and can be realized in mixtures of self-catalytic active colloids, where the local chemistry mediates non-reciprocal interactions among the two species, as demonstrated for instance in  \cite{Saha2019,Agudo-Canalejo2019} via numerical simulations.

\section{\label{sec:md}Continuum model}

We consider a binary mixture described by two conserved phase fields $\phi_{A}$ and $\phi_{B}$ with Cahn-Hillard dynamics~\cite{Cahn1958,Elliott1989,Khain2008} augmented by cross-diffusion \footnote{The natural coupling of two scalar fields with Model B dynamics coming from a term $\sim g\phi_{A}^{2}\phi_{B}^{2}$ in the free energy density is known to yield a rich phase diagram with the possibility of tetracritical points and first order transitions \cite{Liu1973,Wegner1973}, as pointed out to us by David Nelson. We plan to explore the effect of nonreciprocal biquadratic couplings of this type in future work.},
\begin{equation}
 \label{eq:ptPhi}
 \partial_t \phi_{\mu}=\bm\nabla\cdot\left[\left(\chi_\mu+\phi_\mu^2-\gamma_\mu\nabla^2\right)\bm\nabla\phi_\mu+\kappa_{\mu\nu}\bm\nabla\phi_{\nu})\right]\;,
 \end{equation}
where $\mu,\nu=A,B$ and no summation is intended.
In the absence of cross-diffusive couplings ($\kappa_{\mu\nu}=0$),  the fields are decoupled, with ground states   $\phi_{\mu}^g=0$ for $\chi_{\mu}>0$, describing homogeneous states,  and $\phi_\mu^g=\pm \sqrt{-3\chi_{\mu}}$ when $\chi_{\mu}<0$, corresponding to  phase separated states.

The cross-diffusivities  control interspecies interaction, allowing  phase gradient of one species to drive currents of the other species. Equal cross-diffusivities, $\kappa_{AB}=\kappa_{BA}=\kappa$, yield an effective repulsion between the two fields. When sufficiently strong to overcome the entropy of mixing, such a repulsion results in the formation of spatial domains of high/low $\phi_A/\phi_B$, i.e., a demixed state. Here, in contrast, we introduce non-reciprocity by allowing these two quantities to have opposite signs~\footnote{We note that in a binary mixture of diffusing particles, the cross diffusivities would differ as each $\kappa_{\mu\nu}$ would depend on the concentration of the two species as required to maintain detailed balance,  but they would always have  the same sign.}, as can for instance be achieved in mixtures of active and passive Brownian particles (see SI section VI.B) or in mixtures of  colloids with competing repulsive and attractive interactions (see SI section VI.A).
We tune the degree of non-reciprocity $\delta>0$ by letting
\label{eq:kappa}
\begin{align}
\kappa_{AB}=&\kappa-\delta\;, \notag\\
\kappa_{BA}=&\kappa+\delta\;.
\end{align}
As shown below,  this non-reciprocity breaks PT symmetry and gives rise to spatiotemporal patterns  of $\phi_{A}$ and $\phi_{B}$ that break both spatial and temporal translation symmetry.

We have studied numerically \eqref{eq:ptPhi} in a one-dimensional box of length $L=2\pi$, for the case where $\chi_A<0$, $\chi_B>0$, and ignoring $\phi_{B}^{2}$ in the self-diffusivity. The results are easily generalized to the case where both components are supercritical ($\chi_A<0$ and $\chi_B<0$) and to higher dimensions (see SI sections IV-V), but remain qualitatively unchanged. We have integrated ~\eqref{eq:ptPhi}  with a fourth-order central difference on a uniform grid with spacing $h=2\pi/64$. To march in time, we use a second-order, 128-stage Runge-Kutta-Chebyshev scheme with a time step $\Delta t=0.1$~\cite{Hundsdorfer2003, Verwer1990}. All simulations start from nearly uniform phase fields, where weak random fluctuations are added on top of the initial compositions $\phi_{A}^{0}=0$ and $\phi_{B}^{0}=0$. We fix the values of the parameters as: $\gamma_{A}=0.04$, $\gamma_B=0$, $\chi_{B}=0.005$, $\kappa=0.005$, and study how the system dynamics changes with $\chi_{A}$ and $\delta$.

We find three distinct states by varying $\chi_A$ and  $\delta$, as summarized in Fig.~\ref{fig:Fig1}a. When the cross-diffusivities are reciprocal ($\delta=0$), by increasing  $|\chi_A|$ the system undergoes a Hopft bifurcation from a homogeneous state (gray circles) to a demixed state (blue rectangles) where the two fields are spatially modulated with alternating regions of high $\phi_A$/low $\phi_B$ (Figs.~\ref{fig:Fig1}b,c). This state is stabilized by the cubic term in \eqref{eq:ptPhi}, as in conventional Cahn-Hillard models.  Above a critical value of $\delta$, the demixed state undergoes a second bifurcation to a state where the domains of high $\phi_A$/low $\phi_B$ travel at a constant speed (red triangles in Fig.~\ref{fig:Fig1}a, see also Figs.~\ref{fig:Fig1}b,~\ref{fig:Fig1}d for spatiotemporal patterns). The velocity of the traveling pattern provides an order parameter for this transition, and the direction of motion is picked spontaneously. The opposite signs of the cross-diffusivities provide effective antagonistic repulsive and attractive interactions between the two fields.  The drift bifurcation is triggered by the nucleation of a phase shift in the spatial modulation of the two fields that allows species  $A$ to outrun $B$, while $B$ tries to catch up with $A$. At weak non-reciprocity, species $A$ is too slow to escape from $B$, and the static pattern is restored.  Strong non-reciprocity, on other hand, allows species $A$ to outrun $B$. As the distance between the two increases, $A$ gradually slows down while $B$ speeds up until the two share a common speed and become trapped in a steady traveling state. This ``run-and-catch'' scenario is quantified below with a simple one-mode analysis of our dynamical equations that captures the behavior quantitatively. The transitions between the various states obtained from the one-mode approximation are shown as solid lines in Fig.~\ref{fig:Fig1}a and provide an excellent fit to the numerics in one dimension. Finally, as discussed further below, the transition is associated with the breaking of reflection symmetry or parity of the spatial modulations, as well as of time-reversal symmetry, hence provides a realization of a PT-breaking transition.  

We show in the SI that the same scenario applies qualitatively in two dimensions. In this case, in addition to traveling spatial structures, we also observe oscillatory patterns that are absent in 1D. In the oscillatory state the system organizes into high/low concentration region of each species that periodically split and merge. The frequency of oscillation increases with $\delta$, suggesting that the oscillating states are a richer manifestation of non-reciprocity and of the ``run-an-catch'' mechanism that controls the dynamics in 1D. Both traveling and oscillatory states appear to be stable and coexist at high $\delta$, with the state selection being controlled by initial conditions. This suggests  that it would be interesting to go beyond the deterministic model considered here to examine the role of noise. A full study of 2D systems will be reported elsewhere.

\section{One-mode approximation}
To uncover the physics behind the PT-breaking bifurcation,  we expand the fields $\phi_\mu$ in a Fourier series as $ \phi_{\mu}(x,t)=\sum_{j=-\infty}^{\infty}\hat{\phi}_{\mu}^{j}(t)e^{iq_{j}x}$, where $\hat{\phi}_{\mu}^{j}=(2\pi)^{-1}\int_{0}^{2\pi}dx\phi_{\mu}e^{-iq_{j}x}$ is the amplitude of mode $j$. Substituting this in \eqref{eq:ptPhi}, and apply the Galerkin method ~\cite{Hesthaven2007}, one obtains a set of coupled ordinary differential equations for the Fourier amplitudes. For the one dimensional model described above, we have verified numerically that only the first Fourier mode $q_{1}=1$ is activated. We can then replace the original partial differential equations with a single-mode approximation, given by
\begin{subequations}
  \label{eq:dtPhiFC}
\begin{align}
  \frac{d \hat{\phi}_{A}^{1}}{d t}=&-\left(\alpha_{A} +|\hat{\phi}_{A}^{1}|^{2}\right)\hat{\phi}_{A}^{1}-(\kappa-\delta)\hat{\phi}_{B}^{1}, \\
    \frac{d \hat{\phi}_{B}^{1}}{d t}=& -\alpha_{B}\hat{\phi}_{B}^{1} -(\kappa+\delta)\hat{\phi}_{A}^{1},
\end{align}
\end{subequations}
where $\alpha_{A}=\chi_{A}+\gamma_{A}+(\phi_{A}^{0})^{2}$ can be negative and $\alpha_{B}=\chi_{B}>0$. When $\chi_B>0$ the cubic term in the dynamics of $\phi_B$   simply provides a higher order damping  and can be neglected. Writing the complex amplitudes in terms of amplitudes and phases as $\hat{\phi}_{\mu}^{1}=\rho_{\mu}e^{i\theta_{\mu}}$, \eqref{eq:dtPhiFC} can be written as
\begin{subequations}
  \label{eq:dtPhiAP}
\begin{align}
\label{eq:dtPhiAA}
\dot\rho_{A}=&-(\alpha_{A} +\rho_{A}^{2})\rho_{A}-(\kappa-\delta)\rho_{B}\cos\theta\;, \\
\label{eq:dtPhiAB}
\dot\rho_{B}=&-\alpha_{B}\rho_{B}-(\kappa+\delta)\rho_{A}\cos\theta\;, \\
\label{eq:dtTheta}
\dot\theta=&\left[(\kappa-\delta)\rho_{B}/\rho_{A}+ (\kappa+\delta)\rho_{A}/\rho_{B}\right]\sin\theta\;, \\
  \label{eq:dtThetaA}
 \dot\Phi=&\left[(\kappa-\delta)\rho_{B}/\rho_{A}- (\kappa+\delta)\rho_{A}/\rho_{B} \right]\sin\theta\;
\end{align}
\end{subequations}
where $\theta\equiv\theta_{A}-\theta_{B}$ and $\Phi\equiv\theta_{A}+\theta_{B}$ are the difference and sum of the two phases. Note that the sum phase $\Phi$ is slaved to the other quantities. A  broken PT pattern traveling at constant velocity corresponds to  $\dot\rho_A=\dot\rho_B=\dot\theta=0$ and  $\dot\Phi={\rm constant}$, which requires $\sin\theta\not=0$ and $(\kappa-\delta)\rho_{B}/\rho_{A}+ (\kappa+\delta)\rho_{A}/\rho_{B}=0$, or equivalently
$\kappa_{AB}\rho_{B}^{2}=-\kappa_{BA}\rho_{A}^{2}$, hence the two cross-diffusivities must have opposite signs. As we will see below, this is a necessary, but not sufficient condition for the existence of the traveling state. Next, we examine the fixed points of \eqref{eq:dtPhiAA}-\eqref{eq:dtTheta} and their stability.

\paragraph{Fixed points.} There are three fixed points: a trivial fixed point ($F_H$) with $\rho_A=\rho_B=0$ ($\theta$ and $\Phi$ are undetermined), corresponding to a homogeneous mixed state, and two non-trivial fixed points, corresponding to static ($F_S$) and traveling ($F_T$) demixed states. The state $F_S$ describes out-of-phase spatial variations of the two phases, with $  \theta^{s}=\pi$ and
\begin{subequations}
  \label{eq:FS}
  \begin{align}
    \label{eq:rhoAS}
    \rho_{A}^{s}=&\left(\frac{\kappa^{2}-\delta^{2}-\alpha_{A}\alpha_{B}}{\alpha_{B}} \right)^{1/2}, \\
    \label{eq:rhoBS}
    \rho_{B}^{s}=&\left( \kappa+\delta \right)\rho_{A}^{s}/\alpha_{B},
\end{align}
\end{subequations}
while $\Phi$ remains undetermined. This solution of course only exists provided $\alpha_{A}\alpha_{B}<\kappa^{2}-\delta^{2}$. Since $\alpha_{B}>0$, the onset of the static demixed state requires $\alpha_{A}<0$ to drive the growth of $\rho_{A}$, which is then saturated by the cubic damping in  \eqref{eq:dtPhiAA}. Interspecies interactions modulate the pattern, resulting in out-of-phase spatial variations of $\phi_A$ and $\phi_B$, while $\dot\theta_A=\dot\theta_B$ remains zero, i.e., the modulation is static. Note that in this state the two fields, although out of phase, have the same parity, either both even or both odd functions of $x$.

\begin{figure}[t]
\centering
\includegraphics[width=\linewidth]{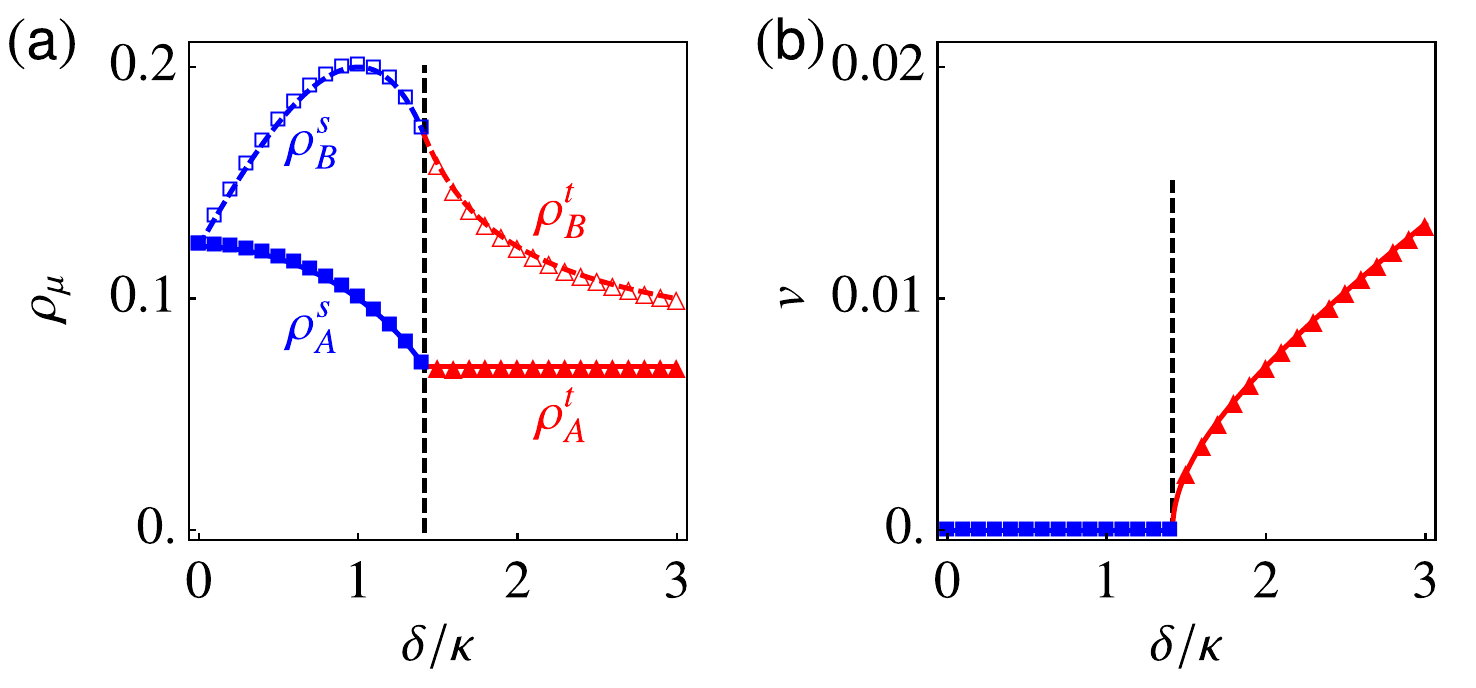}
\caption{\label{fig:Fig2} (a) Comparison of the amplitude of the first Fourier modes as obtained from simulations (symbols) and the one-mode approximations (lines):  $\rho_{A}$ (solid line and filled symbols) and $\rho_{B}$ (dashed line and empty symbols) as functions of $\delta/\kappa$. (b) Velocity of the traveling pattern as a function of  $\delta/\kappa$ from simulations (symbols) and one-mode approximation (line). In both (a) and (b) the black vertical dashed line denotes the critical value $\delta_{c}/\kappa=(1+\alpha_{B}^2/\kappa^{2})^{1/2}$ of the static-to-traveling transition. $\chi_{A}=-0.05$ is used in both panels.}
\end{figure}

The $F_T$ state is a spatial modulation traveling at constant speed
\begin{equation}
\label{eq:speed}
v=\dot\Phi^{t}/2=\pm\sqrt{\delta^{2}-\delta_c^{2}}\sim(\delta-\delta_c)^{1/2}\;,
\end{equation}
with $\delta_c=\sqrt{\kappa^2+\alpha_B^2}$ the critical value of nonreciprocity required for the establishment of the traveling pattern, and
\begin{subequations}
  \label{eq:FT}
  \begin{align}
    \label{eq:rhoAT}
    \rho_{A}^{t}=&\left(-\alpha_{A}-\alpha_{B} \right)^{1/2}, \\
    \label{eq:rhoBT}
    \rho_{B}^{t}=&\sqrt{(\delta+\kappa)/(\delta-\kappa)}~\rho_{A}^{t}, \\
    \label{eq:ThetaT}
  \theta^{t}=&\textrm{arccos}\left( -\sqrt{\frac{\alpha_{B}^{2}}{\delta^{2}-\kappa^{2}}} \right).
\end{align}
\end{subequations}
As we will see below, the speed $v$ provide the order parameter for the transition form the static to the traveling state.
This latter of course only exists when $\kappa-\delta<0$, or more specifically it requires both $\alpha_{A}<-\alpha_{B}$ and $\delta^{2}\ge\kappa^{2} +\alpha_{B}^{2}$, i.e., strong enough non-reciprocity. It arises because a solution with $\sin\theta\not=0$ allows each field to travel at a  finite velocity $v_{\mu}=\dot\theta_\mu$.  The direction of each $v_\mu$ is set by fluctuations or initial conditions.  As shown in Fig.~\ref{fig:Fig2}, the velocity of the traveling modulation and the spatial profiles of the two fields obtained from the one-mode approximation provide an excellent fit to those extracted from numerical solution of Eq. \eqref{eq:ptPhi}.
As discussed below, the traveling pattern breaks the reflection symmetry (parity) of the static one, as well as time reversal invariance.

\paragraph{Linear stability analysis.} A linear stability analysis of the fixed points yields the boundaries between the various states shown in Fig.~\ref{fig:Fig1}a and provides a clear understanding of the mechanism of the drift instability. Linearizing ~\eqref{eq:dtPhiFC} about the homogeneous state reveals that in this state the dynamics of fluctuations is controlled by two eigenvalues given by
\begin{equation}
  \label{eq:lambdaH}
    \lambda_{\pm}=-\frac{1}{2}(\alpha_{A}+\alpha_{B})
      \pm\frac12\sqrt{(\alpha_{A}-\alpha_{B})^{2}+4(\kappa^2-\delta^2)}\;.
\end{equation}
If $\delta^2<\kappa^2+(\alpha_A-\alpha_B)^2/4$, the eigenvalues are real. The largest eigenvalue $\lambda_+$  becomes positive, signaling an instability, when $\delta^{2}=\kappa^{2}-\alpha_{A}\alpha_{B}$. This diffusive instability is displayed as a blue line in Fig.~\ref{fig:Fig1}a.  It is a super-critical pitchfork bifurcation, where the trivial steady state $F_H$ undergoes spontaneous breaking of translational symmetry leading to the transition to the static phase-separated state $F_S$.  Conversely, when $\delta^2>\kappa^2+(\alpha_A-\alpha_B)^2/4$ the eigenvalues are complex conjugate.  The state $F_H$ can still become unstable when $\alpha_{A}<-\alpha_{B}$,  albeit now via an oscillatory instability shown as a red line in Fig.~\ref{fig:Fig1}a.

Further insight is gained by examining the stability of $F_{S}$. This requires the analysis of the eigenvalues of the $3\times 3$ matrix obtained by  linearizing \eqref{eq:dtPhiAA}-\eqref{eq:dtTheta}. Details are given in the SI.  Note that the matrix is block diagonal, coupling separately the two amplitudes and the phase difference $\theta$. One finds that the instability is driven by the growth of fluctuations in the relative phase $\theta$ that become unstable when $\delta>\delta_c$. This boundary $\delta=\delta_c$ corresponds to the appearance of $F_{T}$ and is shown as a black line  in Fig.~\ref{fig:Fig1}a.  The instability of the relative phase is associated with the ``run-and-catch'' scenario described earlier and signals the transition to a state where the two fields have a constants phase lag (different from $\pi$), while traveling with a common speed.

\begin{figure}[t]
\centering
\includegraphics[width=\linewidth]{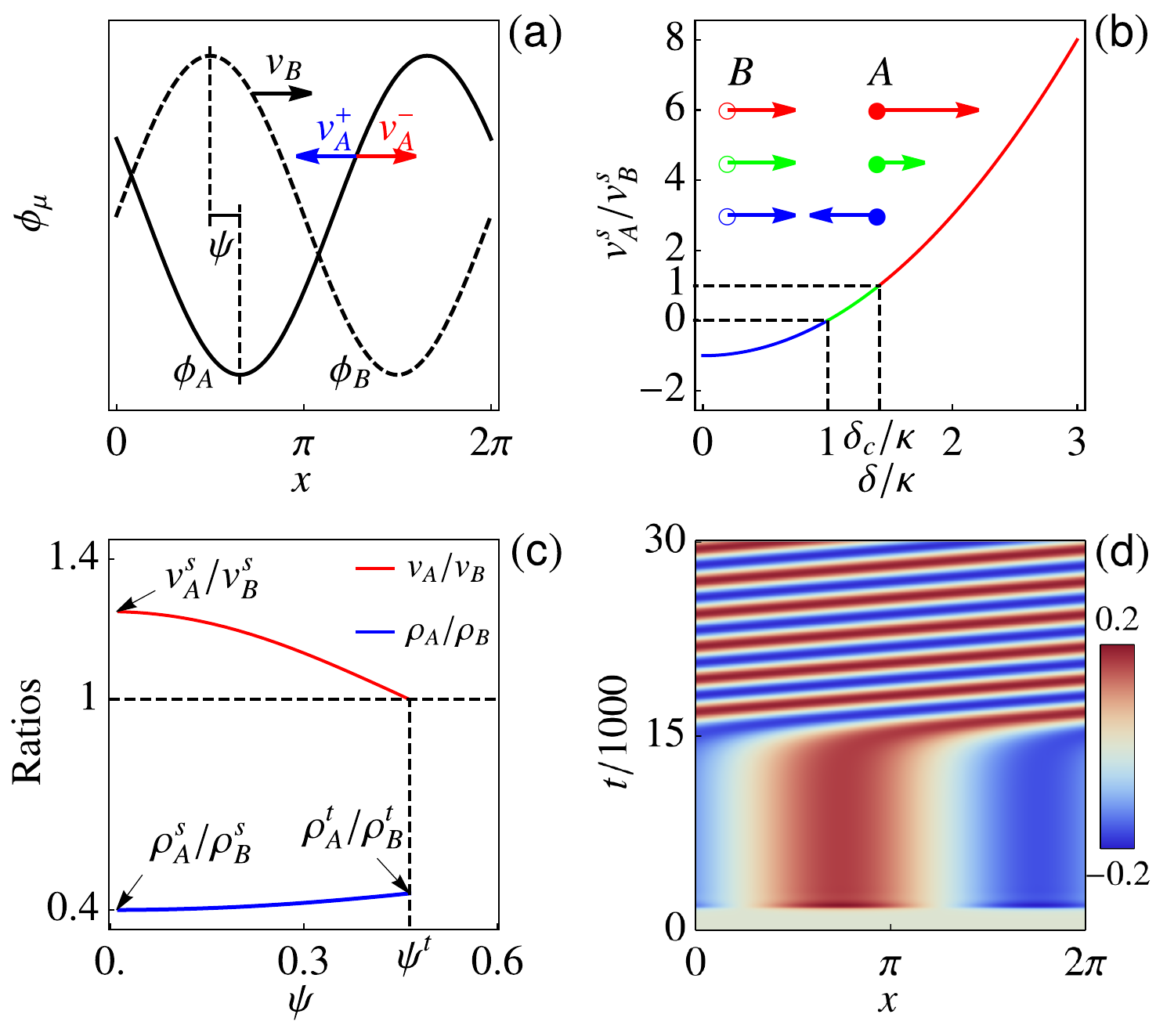}
\caption{\label{fig:Fig3} A pictorial explanation of the ``run-and-catch'' mechanism that leads to the stable traveling state. (a) A phase shift $\psi=\theta-\pi$ of $\phi_{A}$ (solid line) and $\phi_{B}$ (dashed line) relative to the out-of-phase modulation of the stationary demixed state results in finite velocities for the two fields. For a given $v_B$ (black arrow), $v_A$  is in the opposite (blue) or same (red) direction depending on whether the cross diffusivities have equal or opposite signs. (b) Ratio of velocities of the two species as given by \eqref{eq:VRatio} evaluated at the $F_S$ fixed point as a function of $\delta/\kappa$. The arrows are a depiction of the velocities obtained for a small finite $\psi$. Arrow colors correspond to the colors of each portion of the curve. (c) Ratio of the velocities (red) and amplitudes (blue) of the two species as functions of $\psi=\theta-\pi$ obtained from a simulation where the static pattern experiences a drift bifurcation and becomes traveling (i.e. panel d). The values at $\psi=0$ and $\psi=\psi^t=\theta^t-\pi$ agree with those obtained from the one-mode model. The horizontal dashed line corresponds to $v_{A}/v_{B}=1$ and the vertical dashed line is the value $\psi^{t}$ corresponding to the steady traveling state.(d) Spatiotemporal patterns at the onset of the static-to-traveling transition with $\delta/\kappa=1.5$. $\chi_{A}=-0.05$ is used.}
\end{figure}

To highlight the mechanism responsible for the traveling pattern, note that the velocity of the field modulations $v_\mu=\dot\theta_\mu$ are given by $v_A=\kappa_{AB}(\rho_B/\rho_A)\sin\theta$ and  $v_B=-\kappa_{BA}(\rho_A/\rho_B)\sin\theta$, hence are identically zero in the static state $F_S$ where $\theta^{s}=\pi$. Now consider the effect of a small fluctuation in the relative phase by letting $\theta=\pi+\psi$, as shown in Fig.~\ref{fig:Fig3}a. Evaluating the amplitudes at the steady state values, the  velocities are then given by $v_A^{s}=-\frac{\kappa_{AB}\kappa_{BA}}{\alpha_B}\psi$ and $v_B^{s}=\alpha_B\psi$ (see Fig.~\ref{fig:Fig3}b).
If the cross-diffusivities $\kappa_{AB}$ and $\kappa_{BA}$ have the same sign, the two species move in opposite directions (black and blue arrows in Fig. \ref{fig:Fig3}a), exerting reciprocal driving forces on each other, and the perturbation $\psi$ decays.  On the other hand, if $\kappa_{AB}$ and $\kappa_{BA}$ have opposite signs, the two species travel in the same direction (black and red arrows in Fig. \ref{fig:Fig3}a) and can play run-and-catch with each other. To establish the precise condition for the onset of the traveling state, it is useful to examine the ratio of the two velocity, which is well defined even in the static demixed state and is given by
\begin{equation}
\label{eq:VRatio}
\frac{v_{A}}{v_{B}}=-\frac{\kappa_{AB}\rho_{B}^{2}}{\kappa_{BA}\rho_{A}^{2}}=-\frac{(\kappa-\delta)\rho_{B}^{2}}{(\kappa+\delta)\rho_{A}^{2}}\;.
\end{equation}
In the stationary demixed state, where $\rho_B^s/\rho_A^s=(\kappa+\delta)/\alpha_B$, we find $v_A^s/v_B^s=(\delta^2-\kappa^2)/\alpha_B^2$. This quantity is shown in Fig. \ref{fig:Fig3}b. When  $v_A^s/v_B^s<0$ (blue portion of the curve) a small fluctuation $\psi=\theta-\pi$ of the relative phase yields opposite field velocities (blue arrows), while when $v_A^s/v_B^s>0$ the velocities are in the same direction (green portion of the curve and green arrows). Only when  $v_A^s/v_B^s>1$, however, nonreciprocity is strong enough to destabilize the static pattern (red line and arrows in Fig. \ref{fig:Fig3}b). The onset of the traveling state corresponds to $v_A^s=v_B^s$ or $\delta=\delta_c$,  as obtained from the linear stability analysis. The condition $v_A=v_B$ provides a \emph{general } necessary condition for the onset of traveling patterns of two interacting scalar fields.

The equality of the velocities  is not, however, sufficient to stabilize the traveling pattern as the perturbation $\psi$ will keep increasing if $v_{A}>v_{B}$ persists.
Non-reciprocal interactions come again to the rescue by facilitating the ``redistribution'' of amplitude growth.
 Specifically, as $\psi$ increases, both the damping of $\rho_{A}$ and the activation of $\rho_{B}$ originating from the non-reciprocal nature of the cross couplings  become weaker (last terms in \eqref{eq:dtPhiAA}-\eqref{eq:dtPhiAB}). Consequently, the amplitude ratio $\rho_{A}/\rho_{B}$ increases and suppresses the velocity difference until $v_{A}=v_{B}$, allowing the development of a steady traveling pattern, as shown in Fig.~\ref{fig:Fig3}c. We have validated this simple picture displayed in Fig.~\ref{fig:Fig3}d by examining numerically the mechanisms of  stabilization of the traveling state $F_T$ for $\delta$ slightly larger than $\delta_c$.

\section{Static-to-traveling as a PT-breaking transition}

The static-to-traveling transition described in this work belongs to a more generic class of PT-breaking transitions~\cite{Coullet1989, El-ganainy2018, Hanai2019}, which has been studied in optical and quantum systems~\cite{Lin2011, Hanai2019}  and more recently in polar active fluids with non-reciprocal interactions~\cite{Fruchart2020}.  This type of transition is known to occur at a so-called exceptional point, which is simply a point where the eigenvalues of the matrix that governs the linear stability of a fixed point become equal and its eigenvectors are co-linear.    While not uncommon in hydrodynamics when fluids are driven by external forces or in systems described by non-conserved fields, the occurrence of such a transitions giving rise to nontrivial traveling structures in \emph{conserved} systems is unexpected.

The dynamics of our coupled fields can be written in a compact form as
\begin{equation}
\partial_t\begin{pmatrix} \phi_A\\\phi_B\end{pmatrix}=\mathbf{M}\cdot\begin{pmatrix} \phi_A\\\phi_B\end{pmatrix}\;,
\end{equation}
where the $2\times 2$ matrix operator $\mathbf{M}=\mathbf{M}[\phi_A^2,\phi_B^2,\nabla^2]$ can be inferred from \eqref{eq:ptPhi}.  In the static, spatially modulated solution corresponding to the demixed state, the two fields $\phi_A$ and $\phi_B$ are out of phase, but have the same parity under spatial inversion, $x\rightarrow -x$, as required by the symmetry of $\mathbf{M}$. The domains become traveling by acquiring a component of the opposite parity that breaks the relative parity of the two fields, as described in Ref.~\cite{Coullet1989}. Hence the transition to the traveling state breaks both parity and time reversal invariance.

This is most easily understood in the context of the one-mode approximation by considering a static $F_S$ solution of the form $\phi_{B}=\rho_{B}\cos(x)$  and $\phi_{A}=\rho_{A}\cos(x+\pi)$. Both fields are even and are out of phase, but have different amplitudes.  A perturbation $\psi$ in the phase difference yields  $\phi_{A}=\rho_{A}\cos(x+\pi+\psi)=\rho_A\left[-\cos\psi\cos x+\sin\psi\sin x\right]$, breaking parity as  $\phi_{A}$ now acquires an odd component.
The response to such a perturbation is governed by \eqref{eq:dtTheta} linearized about the steady state for $\delta\rightarrow\delta_c$, which is given by
\begin{equation}
\label{eq:dtPsi}
\dot\psi\simeq\frac{2\delta_c(\delta-\delta_c)}{\alpha_B}\psi\;.
\end{equation}
For $\delta<\delta_c$, the odd component of $\phi_A$ proportional to $\psi$ decays, restoring the parity of the static solution.  For $\delta>\delta_c$, $\psi$ grows to a finite value, destabilizing the static state. As a result, $\phi_A$ acquires a finite odd component, breaking the parity of the static solution.  Meanwhile, near the transition \eqref{eq:dtThetaA} gives $\dot\Phi_T\simeq 2\alpha_B\psi$, resulting in a finite $\dot\Phi$ for $\delta>\delta_c$ and breaking time reversal symmetry.

\section{Discussion and outlook}

We have shown that non-reciprocal effective interactions in a minimal model of \emph{conserved} coupled fields with purely diffusive dynamics lead to a PT-breaking transition to traveling spatially modulated states. While the emergence of traveling spatio-temporal patterns is well known in reaction-diffusion, prey-predators and related models, its appearance in the dynamics of conserved fields without external forcing  is surprising. Although the work presented here is limited to a minimal model in one dimension, preliminary results shown in the SI indicate that the same mechanism is at play in two dimensions, as well as in mixtures of active and passive particles and of particles interacting via competing repulsive and attractive interactions, as may be realized in phoretic colloidal mixtures. We speculate therefore that the mechanism described here  through which  non-reciprocal effective couplings grant motility to static spatial modulations
may be a generic property of multispecies systems describes by scalar fields.

The type of static-to-traveling transition described here occurs in Mullins-Sekerka models of crystal growth~\cite{Mullins1964}, Keller-Segel, prey-predator and reaction-diffusion models of population dynamics and general systems described by non-conserved dynamical fields, where it has been referred to as a drift bifurcation~\cite{Coullet1989,Fauve1991,Biktashev2009}. It occurs in these systems when a stationary or standing wave pattern generated through a conventional Hopf bifurcation undergoes a second instability to a traveling state.  The drift bifurcation can be understood using amplitude equations as arising from the antagonistic coupling of at least two leading modes~\cite{Fauve1991}. Here we show that  a similar mechanism can be at play in multispecies systems with dynamics described by two conserved scalar fields coupled by sufficiently strong nonreciprocal interactions. When  sufficiently strong, nonreciprocity leads to an effective antagonistic repulsion/attraction between the two fields, resulting in the run-and-catch mechanism described here that yields a PT-symmetry breaking transition. Our one-mode approximation provides a minimal analytic description of this generic mechanism, where $v=\dot\Phi/2$ serves as the order parameter for the transition.

A scenario similar to the one described here was recently identified in a binary Vicsek model with non-reciprocal interactions~\cite{Fruchart2020}. The mechanisms promoting the onset of a state with broken PT are the same in both models, but the outcomes are distinct due to the different symmetry of the two systems. In Ref.~\cite{Fruchart2020} it is suggested that non-reciprocal interaction in a \emph{polar} system may generically result in macroscopically chiral phases. Here, in contrast, we consider a \emph{scalar} model with conserved dynamics and demonstrate that in this case non-reciprocity generically yields spatially inhomogeneous traveling states through the same type of PT-breaking transition. Together, these works  pave the way to the study of the interplay of non-reciprocity and spontaneously-broken symmetry, suggesting a path to the classification of a new type of PT-breaking transitions.

Understanding and quantifying the role of non-reciprocity in controlling nonequilibrium pattern formation has direct implication to the assembly of chemically interacting colloids, where different particles naturally produce different chemicals mediating nonreciprocal couplings that can induce the type of chasing behavior seen in our work. It also provides a general framework for understanding the nature of wave and oscillatory behavior seen ubiquitously in systems with non-conserved field, from diffusion reaction to prey-predator and population dynamics models. Our predictions can be tested in detailed simulations of active-passive colloidal mixtures or of particles with antagonistic interactions, as well as experiments in mixtures of chemically driven microswimmers.

Our work opens up many new directions of inquiry. Obvious extensions are to higher dimensions where we expect a richer phase diagram and to systems with birth and death processes that select a scale of spatial patterns~\cite{Cates2010}. The exploration of the role of nonreciprocal interactions  in active matter systems with broken orientational symmetry, either polar or nematic, is only beginning~\cite{Fruchart2020} and promises to reveal a rich phenomenology. Chemically mediated or other nonequilibrium couplings can often be time-delayed, which can provide an additional, possibly competing mechanism for the emergence of oscillatory behavior. Finally, an important open problem is understanding how  nonreciprocity arises as an emergent property in systems with microscopic reciprocal interactions, such as active-passive mixtures.

\acknowledgments
MCM thanks Mark Bowick and Vincenzo Vitelli for illuminating discussions. MCM and ZY were primarily supported by the National Science Foundation (NSF) through the Materials Science and Engineering Center at UC Santa Barbara, DMR-1720256 (iSuperSeed), with additional support from DMR-1609208 and PHY-1748958 (KITP). AB was supported in part by MRSEC-1420382 and BSF-2014279. Finally, the work  greatly benefitted form the intellectuatl stimulation provided by the virtual KITP programs during lockdown.

\bibliography{Nonreciprocity}

\end{document}


\title{Supplemental Material to: Nonreciprocity as a generic route to traveling states}

\author{Zhihong You}
\email[Corresponding author:\ ]{youz@ucsb.edu}
\affiliation{Department of Physics, University of California Santa Barbara, Santa Barbara, CA 93106, USA}
\author{Aparna Baskaran}
\affiliation{Martin Fisher school of Physics, Brandeis University, Waltham, MA 02453, USA}
\author{M. Cristina Marchetti}
\email[Corresponding author:\ ]{cmarchetti@ucsb.edu}
\affiliation{Department of Physics, University of California Santa Barbara, Santa Barbara, CA 93106, USA}

\maketitle

\section{One-mode approximation}
In this section, we show the derivation of Eq. (3) and Eq. (4) in the main text using the Fourier-Galerkin method~\cite{Hesthaven2007}.
\subsection{General form of mode equations}
\begin{figure}[t]
\centering
\includegraphics[width=0.6\linewidth]{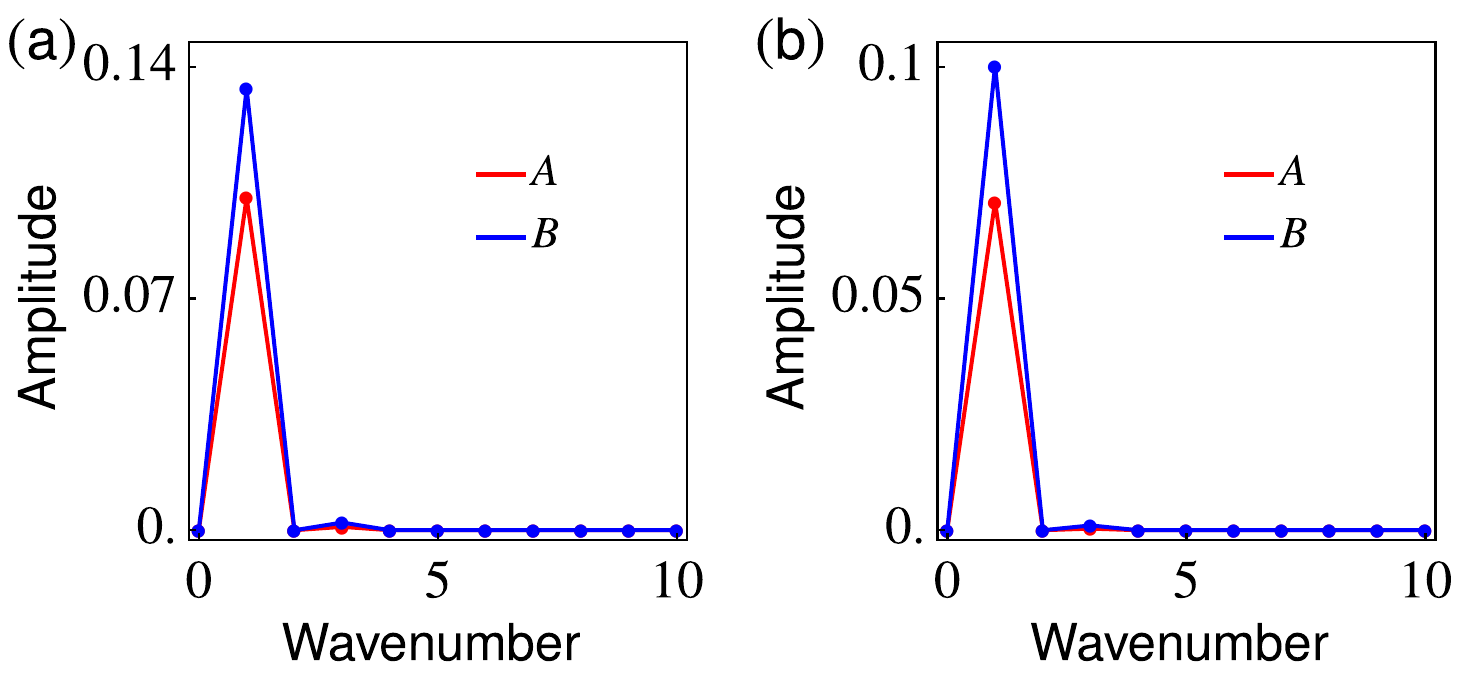}
\caption{\label{fig:Fourier} Amplitudes of Fourier modes $|\hat{\phi}_{\mu}^{j}|$ at the steady (a) static and (b) traveling states. We used $\chi_{A}=-0.05$ and (a) $\delta=\kappa$ and (b) $\delta=3\kappa$.}
\end{figure}

Let us start from the 1D version of Eq. (1),
\begin{subequations}
  \label{eq:ptPhi1}
\begin{align}
\label{eq:ptPhiA1}
\frac{\partial \phi_{A}}{\partial t}=&\partial_{x}\left[\left(\chi_{A}+\phi_{A}^{2}-\gamma_{A}\partial_{x}^{2}\right)\partial_{x}\phi_{A}\right]+\kappa_{AB}\partial_{x}^{2}\phi_{B}, \\
\label{eq:ptRhiB1}
\frac{\partial \phi_{B}}{\partial t}=&\partial_{x}\left(\chi_{B}\partial_{x}\phi_{B}\right)+\kappa_{BA}\partial_{x}^{2}\phi_{A}.
\end{align}
\end{subequations}
Here, we've ignored $\phi_{B}^{2}-\gamma_{B}\partial_{x}^{2}$ as in the main text. To apply the Fourier-Galerkin method, we express $\phi_{\mu}$ as a superposition of Fourier modes:
\begin{equation}
\label{eq:Phi}
  \phi_{\mu}(x,t)=\sum_{j=-\infty}^{\infty}\hat{\phi}_{\mu}^{j}(t)e^{iq_{j}x},
\end{equation}
where the wave number $q_{j}=2\pi j/L$, and
\begin{equation}
\label{eq:PhiF}
  \hat{\phi}_{\mu}^{j}(t)=\frac{1}{L}\int_{0}^{L}\yzh{\phi_{\mu}(x,t)}e^{-iq_{j}x}dx, \\
\end{equation}
are the complex amplitudes of the Fourier modes. The dynamical equations for the Fourier modes obtained from Eq.~\ref{eq:ptPhi1} then are \cite{Hesthaven2007}:
\begin{subequations}
  \label{eq:dtPhiF}
  \begin{align}
    \label{eq:dtPhiFA}
\frac{d\hat{\phi}_{A}^{j}(t)}{dt}=& -q_{j}^{2}\left[\left( \chi_{A}+\gamma_{A}q_{j}^{2} \right)\hat{\phi}_{A}^{j} +\kappa_{AB}\hat{\phi}_{B}^{j} \right]
                                   -\sum_{j_{1},j_{2}}\left(q_{j_{1}}^{2}+2q_{j_{1}}q_{j_{2}}\right)\hat{\phi}_{A}^{j_{1}}\hat{\phi}_{A}^{j_{2}}\hat{\phi}_{A}^{j-j_{1}-j_{2}}, \\
    \label{eq:dtPhiFB}
    \frac{d\hat{\phi}_{B}^{j}(t)}{dt}=& -q_{j}^{2}\left(\chi_{B}\hat{\phi}_{B}^{j}+\kappa_{BA}\hat{\phi}_{A}^{j}\right).
\end{align}
\end{subequations}
By solving for the mode amplitudes $\hat{\phi}_{\mu}^{j}(t)$ and transforming the results back to physical space with Eq. \eqref{eq:Phi}, one can obtain the dynamics of $\phi_{\mu}(x,t)$. This method is usually referred to as the spectral method~\cite{Hesthaven2007}. For the particular system at hand, numerical solutions of the full theory show that the first mode is dominant in determining the dynamical steady states discussed in the main text (Fig. \ref{fig:Fourier}). This is our motivation for considering a reduced theory for our system by truncating the Fourier-Galerkin representation at the level of the first mode.

\subsection{One-mode approximation}

Since $\phi_{A}$ and $\phi_{B}$ are conserved quantities, both $\hat{\phi}_{A}^{0}$ and $\hat{\phi}_{B}^{0}$ are time independent. Setting $\hat{\phi}_{\mu}^{j}=0$ for $|j|> 1$, and $q_{j}=j$, Eq. \ref{eq:dtPhiFA} reduces to
\begin{equation}
\label{eq:dtPhiFA1}
\frac{d\hat{\phi}_{A}^{1}(t)}{dt}=-\left(\alpha_{A}+|\hat{\phi}_{A}^{1}|^{2} \right)\hat{\phi}_{A}^{1} -(\kappa-\delta)\hat{\phi}_{B}^{1},
\end{equation}
where $\alpha_{A}=\chi_{A}+\gamma_{A}+(\phi_{A}^{0})^{2}$. Similarly, for $\hat{\phi}_{B}^{1}$, we have
\begin{equation}
\label{eq:dtPhiFB1}
\frac{d\hat{\phi}_{B}^{1}(t)}{dt}= -\alpha_{B}\hat{\phi}_{B}^{1}-(\kappa+\delta)\hat{\phi}_{A}^{1},
\end{equation}
where $\alpha_{B}=\chi_{B}$. These are Eq. (3) in the main text and constitute the one mode approximation for our full theory.

Expressing the complex quantities $\hat{\phi}_{\mu}^{1}$ in terms of their amplitude and phase, $\hat{\phi}_{\mu}^{1}=\rho_{\mu}e^{i\theta_{\mu}}$, \yzh{Eqs. \eqref{eq:dtPhiFA1} and \eqref{eq:dtPhiFB1}} can be recast as
\begin{subequations}
  \label{eq:dtPhiAP}
\begin{align}
\label{eq:dtPhiAA}
\frac{d \rho_{A}}{d t}=&-(\alpha_{A} +\rho_{A}^{2})\rho_{A}-(\kappa-\delta)\rho_{B}\cos\theta, \\
\label{eq:dtPhiAB}
  \frac{d \rho_{B}}{d t}=&-\alpha_{B}\rho_{B}-(\kappa+\delta)\rho_{A}\cos\theta, \\
\label{eq:dtTheta}
  \frac{d \theta}{d t}=&\left[(\kappa-\delta)\rho_{A}^{-1}\rho_{B}+ (\kappa+\delta)\rho_{A}\rho_{B}^{-1} \right]\sin\theta, \\
  \label{eq:dtPhi}
  \frac{d \Phi}{d t}=&\left[(\kappa-\delta)\rho_{A}^{-1}\rho_{B}- (\kappa+\delta)\rho_{A}\rho_{B}^{-1} \right]\sin\theta,
\end{align}
\end{subequations}
with $\Phi=\theta_{A}+\theta_{B}$ and $\theta=\theta_{A}-\theta_{B}$. These are Eq.~(4) in the main text.

\section{Steady states and their stability}
In this section, we identify the steady states in the one-mode approximation, and study their stability.

\paragraph{Steady states.} The steady states are obtained as the fixed points of \yzh{Eqs. \eqref{eq:dtPhiAA}-\eqref{eq:dtTheta}. Note that $\Phi$ is slaved by the other three quantities.} The solution $\rho_{A}=0$ and $\rho_{B}=0$ corresponds to the homogeneous state, and is referred to as the trivial fixed point $F_{H}$ in the main text. In this case, the phases are  not well defined.

A second fixed point is obtained when $\sin\theta=0$ or $\theta=0,\pi$. Since Eq. \eqref{eq:dtPhiAB} yields $\rho_{B}=-\alpha_{B}^{-1}(\kappa+\delta)\rho_{A}\cos\theta$ and the amplitudes $\rho_\mu$ must be positive  the only acceptable solution is $\theta=\pi$. The second fixed point, $F_{S}$, is then given by
\begin{subequations}
  \label{eq:FS}
  \begin{align}
    \label{eq:rhoAS}
    \rho_{A}^{s}=&\left[(\kappa^{2}-\delta^{2})/\alpha_{B}-\alpha_{A} \right]^{1/2}, \\
    \label{eq:rhoBS}
    \rho_{B}^{s}=&(\kappa+\delta)\rho_{A}^{s}/\alpha_{B}, \\
    \label{eq:ThetaS}
  \theta^{s}=&\pi,
\end{align}
\end{subequations}
In order for $F_{S}$ to exist as a physical state of the system, the argument of the square root in Eq. \eqref{eq:rhoAS} must be real, which requires $\kappa^{2}-\delta^{2}>\alpha_{B}\alpha_{A}$, as given in the main text.

Finally, when $\sin\theta\not=0$ one can have a third solution $F_{T}$ with $\dot\Phi$ finite, corresponding to spatial modulations that travel at the fixed velocity $v=\dot\Phi/2$. This requires $\left[(\kappa-\delta)\rho_{A}^{-1}\rho_{B}+ (\kappa+\delta)\rho_{A}\rho_{B}^{-1} \right]=0$ and it is given by
\begin{subequations}
  \label{eq:FT}
  \begin{align}
    \label{eq:rhoAT}
    \rho_{A}^{t}=&\left(-\alpha_{A}-\alpha_{B} \right)^{1/2}, \\
    \label{eq:rhoBT}
    \rho_{B}^{t}=&\sqrt{(\delta+\kappa)/(\delta-\kappa)}~\rho_{A}^{t}, \\
    \label{eq:ThetaT}
  \theta^{t}=&\textrm{arccos}\left( -\sqrt{\frac{\alpha_{B}^{2}}{\delta^{2}-\kappa^{2}}} \right).
\end{align}
\end{subequations}
The requirement that the arguments of all square roots be positive and the argument of $\textrm{arccos}(\cdot)$  have an absolute value no larger than $1$ yields the conditions of existence for $F_{T}$ as
\begin{subequations}
  \label{eq:ExisFT}
  \begin{align}
    &\alpha_{A}<-\alpha_{B}, \\
    &\delta^{2}-\kappa^{2}\ge \alpha_{B}^{2}.
    \end{align}
\end{subequations}
As we will see below, the first one describes the instability of the stationary spatial modulation $F_S$, while the second one demands that nonreciprocity to be strong enough for the traveling pattern to appear.

\begin{figure}[b]
\centering
\includegraphics[width=0.4\linewidth]{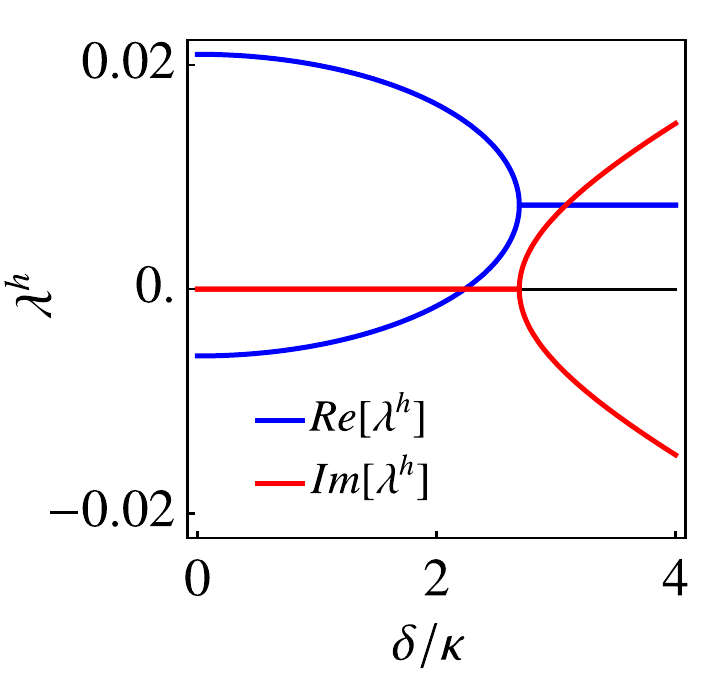}
\caption{\label{fig:lambda} Real (blue lines) and imaginary (red lines) parts of eigenvalues $\lambda^{h}_{\pm}$ as functions of $\delta$ for $\chi_{A}=-0.06$ showing the change from diffusive to propagating modes. An instability is signaled by $Re[\lambda]>0$. }
\end{figure}

\paragraph{Linear stability.} Now we examine the stability of the fixed points, which is controlled by the eigenvalues of the Jacobian matrix evaluated at the fixed points:
\begin{equation}
  \label{eq:M}
  \textbf{M}\equiv
\begin{bmatrix}
\frac{\partial \dot{\rho}_{A}}{\partial \rho_{A}} & \frac{\partial \dot{\rho}_{A}}{\partial \rho_{B}} & \frac{\partial \dot{\rho}_{A}}{\partial \theta} \\
\frac{\partial \dot{\rho}_{B}}{\partial \rho_{A}} & \frac{\partial \dot{\rho}_{B}}{\partial \rho_{B}} & \frac{\partial \dot{\rho}_{B}}{\partial \theta} \\
\frac{\partial \dot{\theta}}{\partial \rho_{A}} & \frac{\partial \dot{\theta}}{\partial \rho_{B}} & \frac{\partial \dot{\theta}}{\partial \theta} \\
\end{bmatrix}
=
\begin{bmatrix}
-\alpha_{A}-3\rho_{A}^{2} & -(\kappa-\delta)\cos\theta & (\kappa-\delta)\rho_{B}\sin\theta \\
-(\kappa+\delta)\cos\theta & -\alpha_{B} & (\kappa+\delta)\rho_{A}\sin\theta \\
\left( \frac{\kappa+\delta}{\rho_{B}}-\frac{(\kappa-\delta)\rho_{B}}{\rho_{A}^{2}} \right)\sin\theta & \left( \frac{\kappa-\delta}{\rho_{A}} -\frac{(\kappa+\delta)\rho_{A}}{\rho_{B}^{2}}\right)\sin\theta &\left( \frac{(\kappa-\delta)\rho_{B}}{\rho_{A}} +\frac{(\kappa+\delta)\rho_{A}}{\rho_{B}}\right)\cos\theta  \\
\end{bmatrix}.
\end{equation}
With our convention, an instability corresponds to a positive real part of the largest eigenvalue.

At the trivial fixed point, $F_{H}$, the phases are undetermined and one can simply examine the stability of Eqs. \eqref{eq:dtPhiFA1} and \eqref{eq:dtPhiFB1}. The corresponding Jacobian matrix at $F_{H}$ is
\begin{equation}
\label{eq:M2}
\textbf{M}_H=
\begin{bmatrix}
-\alpha_{A} & -\kappa_{AB} \\
-\kappa_{BA} & -\alpha_{B} \\
\end{bmatrix},
\end{equation}
with eigenvalues
\begin{equation}
  \label{eq:lambdaH}
  \begin{split}
    \lambda^{h}_{\pm}=\frac{1}{2}\left[-(\alpha_{A}+\alpha_{B})\pm\sqrt{(\alpha_{A}-\alpha_{B})^{2}+4(\kappa^{2}-\delta^{2})}   \right].
    \end{split}
\end{equation}
As discussed in the main text,  the trivial fixed point $F_{H}$ can become unstable through different routes, depending on the sign of the argument of the square root in Eq. \eqref{eq:lambdaH}. When $(\alpha_{A}-\alpha_{B})^{2}+4(\kappa^{2}-\delta^{2})>0$, \yzh{$\lambda^{h}_{\pm}$} are real (Fig. \ref{fig:lambda}), and the homogeneous state becomes unstable via a diffusive instability when $\lambda^{h}_{+}$ becomes positive for
$\kappa^{2}-\delta^{2}>\alpha_{A}\alpha_{B}$. On the other hand, if $(\alpha_{A}-\alpha_{B})^{2}+4(\kappa^{2}-\delta^{2})<0$, $\lambda^{h}_{\pm}$ are complex conjugate (Fig. \ref{fig:lambda}), and the system becomes unstable via an oscillatory instability when $Re[\lambda^{h}_{\pm}]$ becomes positive for $\alpha_{A}<-\alpha_{B}$.

To investigate the stability of $F_{S}$, we evaluate the Jacobian matrix given in Eq. \eqref{eq:M} at the fixed point, with the result
\begin{equation}
  \label{eq:MS}
  \textbf{M}_S
=
\begin{bmatrix}
2\alpha_{A}-3\frac{\kappa^2-\delta^2}{\alpha_B} & \kappa-\delta & 0 \\
\kappa+\delta& -\alpha_{B} & 0 \\
0 & 0 &-\alpha_B-\frac{\kappa^2-\delta^2}{\alpha_B}   \\
\end{bmatrix}
\end{equation}
Fluctuations in $\rho_A$ and $\rho_B$ are coupled. Their relaxation is controlled by the eigenvalues
\yzh{
\begin{equation}
  \label{eq:lambdaS}
    \lambda^{s}_{\pm}=\frac{1}{2}\left[2\alpha_{A}-\alpha_{B}-3\frac{\kappa^2-\delta^2}{\alpha_B} \right]\pm\frac12\sqrt{\left[2\alpha_{A}+\alpha_{B}-3\frac{\kappa^2-\delta^2}{\alpha_B} \right]^2+4(\kappa^2-\delta^2) }
\end{equation}
}
which always have a negative real part in the region of parameters where $F_S$ exists, hence are stable. On the other hand, fluctuations in the relative phase $\theta$ are controlled by
\begin{equation}
    \label{eq:rhoBT}
    \lambda^s_\theta=-\frac{\delta_c^2-\delta^2}{\alpha_{B}}\;,
\end{equation}
where $\delta_c=\sqrt{\kappa^2+\alpha_B^2}$. The eigenvalue is real and positive for $\delta>\delta_c$, signaling the growth of phase fluctuations that destabilize the stationary demixed state, giving rise to traveling patterns. Finally, the eigenvalues of $\textbf{M}$ at $F_{T}$ are too complicated to be instructive. It can, however, be demonstrated numerically, that they are always negative, so the traveling patterns corresponding to this fixed point are always stable, in the region of parameters where it exists.

\section{Critical point as an exceptional point}
As mentioned in the main text, the drift bifurcation induced by the nonreciprocal interspecies interaction belongs to a generic class of phase transitions which has been  studied in optical and quantum systems \cite{Lin2011, El-ganainy2018, Hanai2019}, and more recently in nonreciprocally interacting polar active fluids~\cite{Fruchart2020}. This type of phase transitions is known to occur at a so-called exceptional point that separates a PT-symmetric phase from a phase with broken PT symmetry~\cite{El-ganainy2018,Fruchart2020}.

To clarify these concepts, let us return to the compact form of our one-mode model written in terms of complex Fourier amplitudes, as given in Eq. (3) of the main text,
\begin{equation}
\label{eq:dtPhiM}
\frac{d}{dt}
\begin{bmatrix}
\hat{\phi}_{A} \\
\hat{\phi}_{B}
\end{bmatrix}=
\begin{bmatrix}
  -\alpha_{A}-|\hat{\phi}_{A}|^{2}& -(\kappa-\delta) \\
 -(\kappa+\delta) & -\alpha_{B}
\end{bmatrix}
\begin{bmatrix}
\hat{\phi}_{A} \\
\hat{\phi}_{B}
\end{bmatrix}
\equiv \mathbf{D}[\hat{\phi}_A]\begin{bmatrix}
\hat{\phi}_{A} \\
\hat{\phi}_{B}
\end{bmatrix}\;,
\end{equation}
where we have dropped the mode superscript $1$ to simplify the notation.  The $2\times 2$ non-Hermitian matrix $\mathbf{D}[\hat{\phi}_A]$ in Eq. \eqref{eq:dtPhiM} controls the system's dynamics. The stationary fixed point \yzh{$F_S$} is obtained by solving  $\mathbf{D}[\hat{\phi}_A]=0$
and it is given by 
\begin{equation}
\label{eq:PhiSM}
\textbf{u}^{s}\equiv
\begin{bmatrix}
  \hat{\phi}_{A}^{s} \\
  \hat{\phi}_{B}^{s}
  \end{bmatrix}
=
\begin{bmatrix} \sqrt{(\kappa^{2}-\delta^{2})/\alpha_{B}-\alpha_{A}}~e^{i\theta_{0}} \\
-\alpha_{B}^{-1}(\kappa+\delta)\hat{\phi}_{A}^{s}
  \end{bmatrix},
\end{equation}
where $\theta_{0}$ is an undetermined phase indicating translation of $\phi_{\mu}$ in space. Since $\hat{\phi}_{B}^{s}\sim -\hat{\phi}_{A}^{s}$, the two complex fields are out of phase and this fixed point corresponds exactly to the out-of-phase static patterns described in the main text.

Let us now assume that there exists a stationary nontrivial solution that solves  $\mathbf{D}[\hat{\phi}_A^0]\begin{bmatrix}
\hat{\phi}_{A}^0 \\
\hat{\phi}_{B}^0
\end{bmatrix}=0$ and 
 evaluate eigenvalues and eigenvectors of $\mathbf{D}_0\equiv\mathbf{D}[\hat{\phi}_A^0]$ and then examine their behavior at the stationary fixed point given in \eqref{eq:PhiSM}.
These are given by
\begin{equation}
\label{eq:lambdaM}
\lambda_{\pm}=-\frac{1}{2}\left(\alpha_B+\alpha_A+|\hat{\phi}_A^0|^2\right)\pm\frac12\sqrt{\Delta}\;,
\end{equation}
where $\Delta=\left(\alpha_B-\alpha_A-|\hat{\phi}_A^0|^2\right)^2 +4(\kappa^2-\delta^2)$ and 
\begin{equation}
\label{eq:uM}
\textbf{u}_{\pm}=C_\pm
\begin{bmatrix}
-(\lambda_\pm+\alpha_B) \\
\kappa+\delta
\end{bmatrix}
\end{equation}
where $C_\pm$ are normalization constants.
If $\Delta>0$, the eigenvalues are real. If stable, the solutions will be stationary in this regime. Note the matrix  $\mathbf{D}_0$ is in general different form the linear matrix that controls the stability of fixed points. 
If $\Delta<0$, the eigenvalues are complex conjugate, signaling the onset of an oscillatory solutions. The point $\Delta=0$ corresponds to 
$\delta^2=\kappa^2+(\alpha_B-\alpha_A-|\hat{\phi}_A^0|^2)^2/4$. Substituting $\phi_A^0=\phi_A^s$, we find that the modes change from real to complex conjugates at $\delta=\delta_c=\sqrt{\kappa^2+\alpha_B^2}$.
At this point, solutions change from stationary to oscillatory.

We also note that when evaluated at the fixed point $F_S$ the eigenvalues become
\begin{equation}
\label{eq:lambdaMs}
\lambda_{\pm}=-\frac{1}{2\alpha_B}\left((\delta_c^2-\delta^2)\pm|\delta_c^2-\delta^2|\right)\;,
\end{equation}
or
\begin{align}
\label{eq:lambda1}
\lambda_1^s=&0\;,\\
\label{eq:lambda2}
\lambda_2^s=&-\frac{\delta_c^2-\delta^2}{\alpha_B}
\end{align}
At the critical point $\delta=\delta_c$, the two eigenvalues are equal and the eigenvectors become co-linear. In addition, the solutions change from stationary to oscillatory. This is what is referred to as an exceptional point~\cite{Kato1966}. This behavior is also associated with the fact that the matrix $\mathbf{D}[\hat{\phi}_A]$ evaluated at the steady state  \eqref{eq:PhiSM} is given by
\begin{equation}
\mathbf{D}_s\equiv \mathbf{D}[\hat{\phi}_A^s]
=
\begin{bmatrix}
  -(\kappa^2-\delta^2)/\alpha_B& -(\kappa-\delta) \\
 -(\kappa+\delta) &  -\alpha_{B}
\end{bmatrix}\;
\end{equation}
hence $\det[\mathbf{D}_s]=0$.

More precisely, to elucidate the nature of the drift instability as an exceptional point, we examine the matrix that controls the linear stability of the stationary density modulated state $F_S$.  We show below that the instability occurs when one of the eigenmodes of such a matrix coalesces with the Goldstone mode associated with spontaneously broken translational symmetry of the modulated state~\cite{Hanai2019, Fruchart2020}. 
To demonstrate this, we linearize  \eqref{eq:dtPhiM}  about the static fixed point $F_{S}$ to obtain coupled equations for the complex fluctuations $\delta\hat{\phi}_\mu$ of the phase fields, given by
\begin{equation}
\label{eq:modeL4}
\frac{d\delta\textbf{u}}{dt}=\textbf{L}\cdot\delta\textbf{u},
\end{equation}
where
\begin{equation}
\label{eq:du}
\delta\textbf{u}=\begin{pmatrix}
  \delta\hat{\phi}_{A} \\
  \delta\hat{\phi}_{B} \\
  \delta\hat{\phi}_{A}^{*} \\
  \delta\hat{\phi}_{B}^{*} 
  \end{pmatrix}
\end{equation}
and
\begin{equation}
\label{eq:du}
\textbf{L}=\begin{pmatrix}
  -\alpha_{A}-2|\hat{\phi}_{A}^{s}|^{2} &-(\kappa-\delta) &-(\hat{\phi}_{A}^{s})^{2} &0 \\
  -(\kappa+\delta) &-\alpha_{B} &0 &0 \\
  -(\hat{\phi}_{A}^{s*})^{2} &0 &-\alpha_{A}-2|\hat{\phi}_{A}^{s}|^{2} &-(\kappa-\delta) \\
  0 &0 &-(\kappa+\delta) &-\alpha_{B}
  \end{pmatrix}
\end{equation}
is the matrix controlling the linear stability of the fixed point. We note that Eqs. \eqref{eq:dtPhiM} are invariant under an arbitrary global translation $\hat{\phi}_{\mu}\rightarrow\hat{\phi}_{\mu}e^{i\delta\theta}$. The breaking of translational symmetry associated with the emergence of the spatially-modulated de-mixed stationary state from the homogeneous state is accompanied by the appearance of a zero mode associated with fluctuations in the total phase of the complex field amplitudes which is the Goldstone mode of the transition. The second PT-breaking transition to the traveling state occurs when another eigenmode of $\textbf{L}$ coalesces with the Goldstone mode at the exceptional point.

\begin{figure}[t]
\centering
\includegraphics[width=1.0\linewidth]{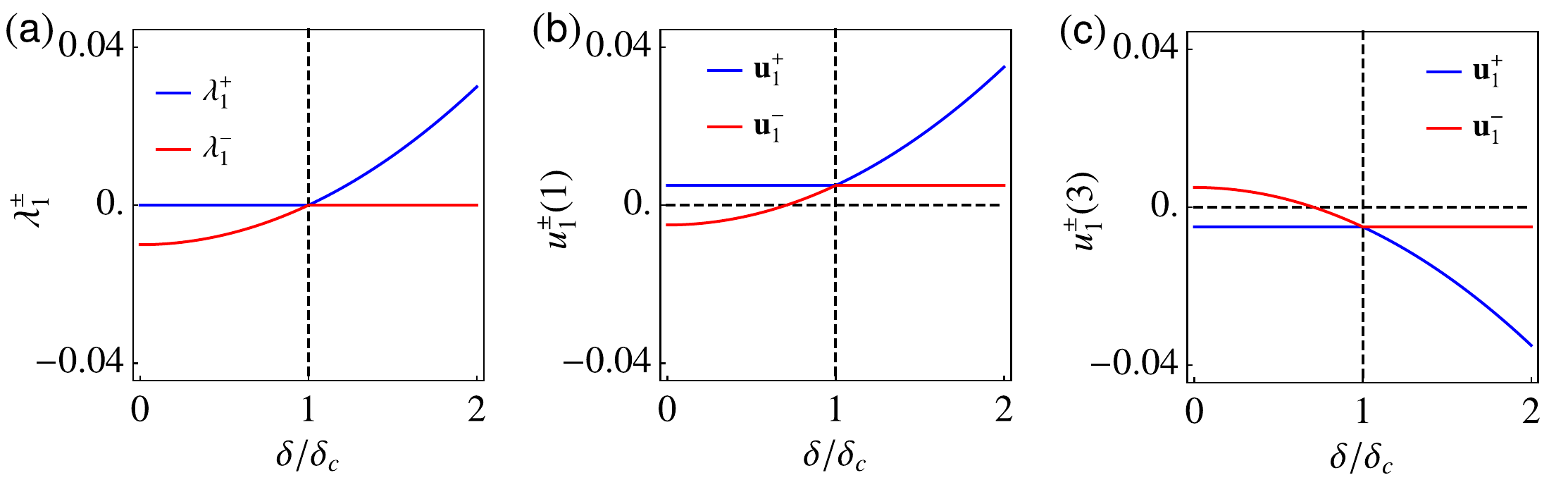}
\caption{\label{fig:EP} (a) Eigenvalues and the (b) first and (c) third elements of eigenvectors of the matrix $\textbf{L}$ evaluated at the static modulated state $F_{S}$. The blue lines corresponds to the Goldstone mode $\textbf{u}_{1}^{+}$, while the red line shows the mode $\textbf{u}_{1}^{-}$, which coalesces with the Goldstone mode at $\delta=\delta_{c}$.}
\end{figure}

The linear matrix $\textbf{L}$ has four eigenvalues
\begin{subequations}
\label{eq:lambL}
\begin{align}
  \lambda_{1}^{\pm}=&\frac{1}{2}\left[ -\alpha_{B}-(\kappa^{2}-\delta^{2})/\alpha_{B}\pm\sqrt{\Delta_{1}} \right]\\
  \lambda_{2}^{\pm}=&\frac{1}{2}\left[ -\alpha_{B}+2\alpha_{A}-3(\kappa^{2}-\delta^{2})/\alpha_{B}\pm\sqrt{\Delta_{2}} \right] 
  \end{align}
\end{subequations}
with the corresponding eigenmodes
\begin{equation}
\label{eq:uL}
  \textbf{u}_{1}^{\pm}=\begin{pmatrix}
  \frac{1}{2}\left[ \alpha_{B}-(\kappa^{2}-\delta^{2})/\alpha_{B}\pm\sqrt{\Delta_{1}} \right] \\
  -(\kappa+\delta) \\
  -\frac{1}{2}\left[ \alpha_{B}-(\kappa^{2}-\delta^{2})/\alpha_{B}\pm\sqrt{\Delta_{1}} \right] \\
  (\kappa+\delta) 
\end{pmatrix},\ 
  \textbf{u}_{2}^{\pm}=\begin{pmatrix}
  -\frac{1}{2}\left[ \alpha_{B}+2\alpha_{A}-3(\kappa^{2}-\delta^{2})/\alpha_{B}\pm\sqrt{\Delta_{2}} \right] \\
  (\kappa+\delta) \\
  -\frac{1}{2}\left[ \alpha_{B}+2\alpha_{A}-3(\kappa^{2}-\delta^{2})/\alpha_{B}\pm\sqrt{\Delta_{2}} \right] \\
  (\kappa+\delta) 
\end{pmatrix},
\end{equation}
where
\begin{subequations}
\label{eq:delt12}
\begin{align}
  \Delta_{1}=&\left[\alpha_{B}+(\kappa^{2}-\delta^{2})/\alpha_{B}\right]^{2}\\
  \Delta_{2}=&\left[ \alpha_{B}+2\alpha_{A}-3(\kappa^{2}-\delta^{2})/\alpha_{B} \right]^{2} +4 \left( \kappa^{2}-\delta^{2} \right)
  \end{align}
\end{subequations}
For $\delta<\delta_{c}$, $\lambda_{1}^{+}=0$ (blue line in Fig. \ref{fig:EP}a), hence $\textbf{u}_{1}^{+}$ corresponds to the Goldstone mode arising from the spontaneous breaking of translational symmetry. At $\delta=\delta_c$, $\Delta_{1}=0$ and $\lambda_1^-$ vanishes and $\textbf{u}_{1}^{-}$ becomes colinear with the eigenvector $\textbf{u}_{1}^{+}$ of the Goldstone mode (Fig.~\ref{fig:EP}b--\ref{fig:EP}c), giving rise to the PT breaking transition or  drift bifurcation, which corresponds to an exceptional point \cite{Hanai2019, Fruchart2020}.

\section{Two-dimensional case}
\begin{figure}[t]
\centering
\includegraphics[width=1.0\linewidth]{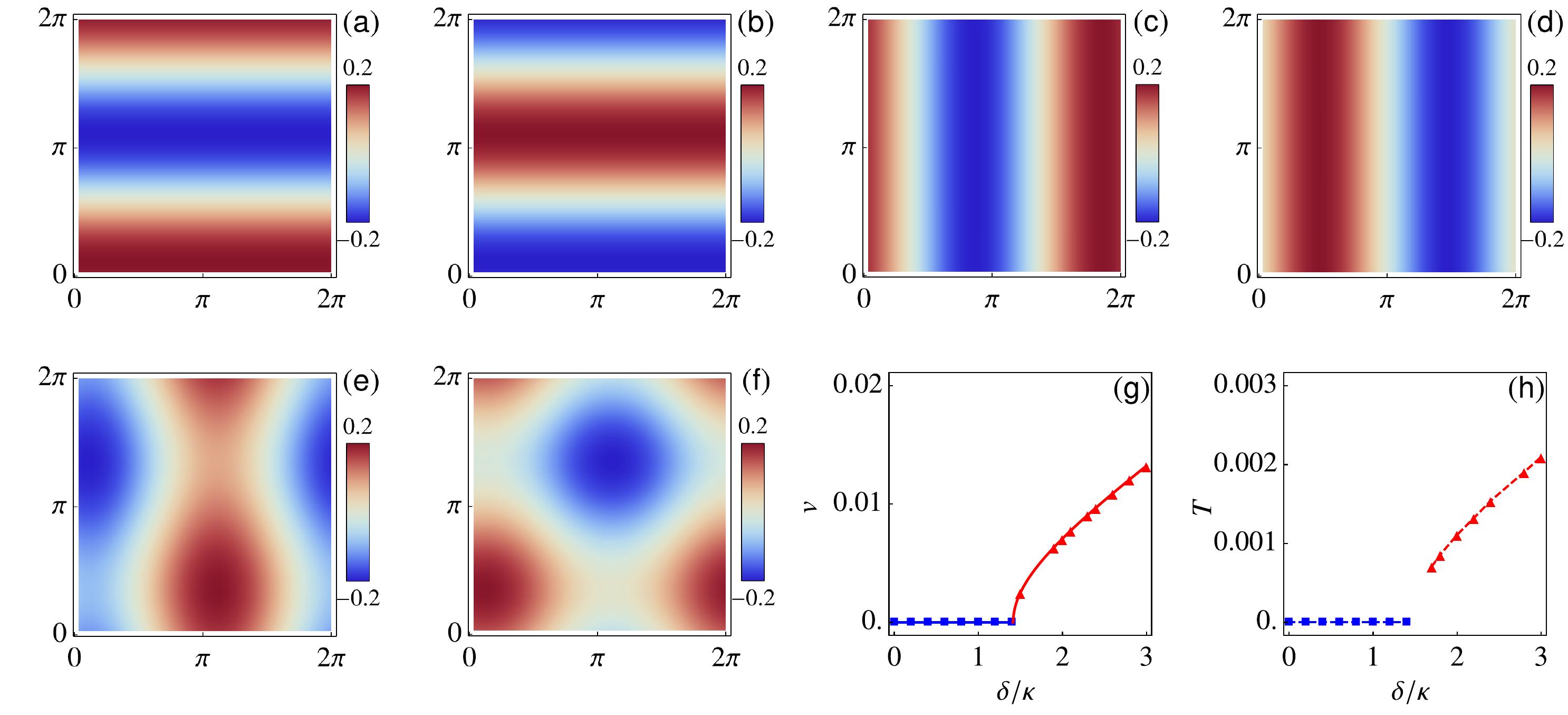}
\caption{\label{fig:2D} Pattern formations in two-dimensional systems at the (a--b) \yzh{static}, (c--d) traveling, and (e--f) oscillatory states. Panels (a,c,e) show $\phi_{A}$, while (b,d,f) $\phi_{B}$, highlighting the phase relation of the two fields. In the stationary demixed state (a--b, $\delta=\kappa$), $\phi_{A}$ and $\phi_{B}$ are out-of-phase, while in the traveling state (c--d, $\delta=3\kappa$) $\phi_{A}$ and $\phi_{B}$ have a constant phase shift different from $\pi$, and the pattern is traveling from left to right. In the oscillatory state(e--f, $\delta=3\kappa$), the high concentration regions of both $\phi_{A}$ and $\phi_{B}$ periodically split and merge. Panels (g) and (h) shows the travellng speed and oscillating frequency of the traveling and oscillatory patterns, respectively. Both increases monotonically with the nonreciprocity $\delta$. In panel (g), the solid line corresponds to $v=\sqrt{\delta^{2}-\delta_c^{2}}$, with $\delta_{c}=\sqrt{\kappa^{2}+\alpha_{B}^{2}}$.}
\end{figure}

We show here that a static-to-traveling transition driven by nonreciprocal interactions also occurs in two dimensional systems, albeit with a richer dynamics. The detailed analysis is left for future work. The goal of this section is to highlight that the qualitatve behavior remains the same. We have integrated numerically  Eq. (1) from the main text in a two-dimensional periodic box of size $L\times L$, with $L=2\pi$ and the same parameter values as in the main text: $\chi_{A}=-0.05$, $\chi_{B}=0.005$, $\gamma_{A}=0.04$, \yzh{$\gamma_{B}=0$}, $\kappa=0.005$, $\phi_{A}^{0}=\phi_{B}^{0}=0$. We have then examined the effect of nonreciprocity by increasing $\delta$.

As in one dimension, we observe  a transition from static to traveling modulations with increasing nonreciprocity, as shown in Figs. \ref{fig:2D}a--\ref{fig:2D}d \yzh{and Movie S1-S2}. The orientation of the band is determined by random fluctuations and by the initial concentration. The expression for drift velocity obtained in one dimension, $v=\pm\sqrt{\delta^{2}-\delta_c^{2}}$, with $\delta_{c}=\sqrt{\kappa^{2}+\alpha_{B}^{2}}$, gives an excellent parameter-free fit to the velocity of the traveling pattern in 2D, as shown in Fig. \ref{fig:2D}g.
In addition to traveling modulations of the fields, in 2D we also observe oscillatory patterns that are absent in the 1D system. This consists of high/low concentration region of each species that periodically split and merge (Figs. \ref{fig:2D}e--\ref{fig:2D}f and \yzh{and Movie S3}). The oscillatory state also originates from the nonreciprocal interactions, as evidenced by the fact that it only appears when $\delta>\delta_c$, and the oscillating frequency increases with nonreciprocity (Fig. \ref{fig:2D}h).  In the parameter region we have explored, the traveling and oscillatory states are both stable, and the state is selected by the initial condition.

\section{Bi-supercritical systems}

\begin{figure}[t]
\centering
\includegraphics[width=0.9\linewidth]{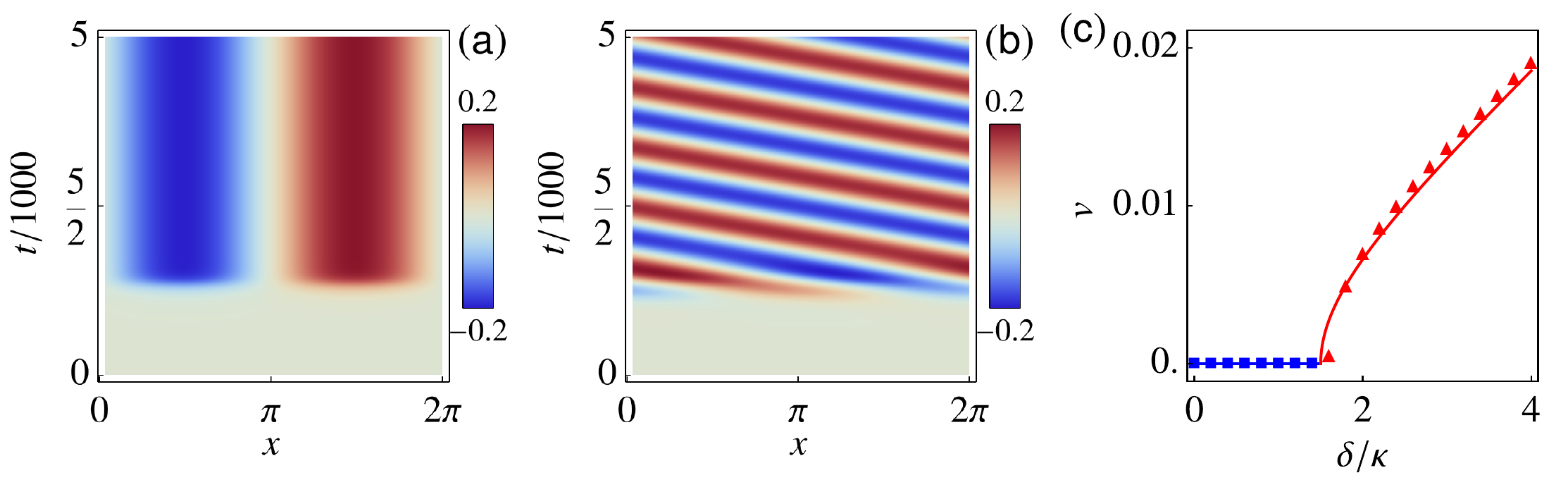}
\caption{\label{fig:2SupeCric} Drift instability in bi-supercritical systems. (a--b) \yzh{Spatiotemporal} patterns of $\phi_{A}$ in (a) static ($\delta=\kappa$) and (b) traveling states ($\delta=2\kappa$). (c) Speed of the traveling modulation as a function of nonreciprocity $\delta/\kappa$.  The red line is a fit to the mean-field theory with $v=\sqrt{\delta^{2}-\delta_c^{2}}$, where $\delta_{c}=1.5\kappa$ is extracted from the best fit to the data. All results shown are for $\chi_{A}=\chi_{B}=-0.05$, $\gamma_{A}=\gamma_{B}=0.04$, $\kappa=0.005$, and $L=2\pi$. }
\end{figure}

In the main text, we have considered the case when field $A$ is supercritical, but field $B$ is subcritical and its dynamics is purely relaxational. To demonstrate the generality of our findings, we have also considered  the case where both fields are supercritical, as obtained when $\chi_A<0$ and $\chi_B<0$. In this case, restoring the term $\phi_{B}^{2}-\gamma_{B}\partial_{x}^{2}$, the equations become
\begin{subequations}
  \label{eq:ptPhi2S}
\begin{align}
\label{eq:ptPhiA2S}
\frac{\partial \phi_{A}}{\partial t}=&\partial_{x}\left[\left(\chi_{A}+\phi_{A}^{2}-\gamma_{A}\partial_{x}^{2}\right)\partial_{x}\phi_{A}\right]+\kappa_{AB}\partial_{x}^{2}\phi_{B}, \\
\label{eq:ptRhiB2S}
\frac{\partial \phi_{B}}{\partial t}=&\partial_{x}\left[\left(\chi_{B}+\phi_{B}^{2}-\gamma_{B}\partial_{x}^{2}\right)\partial_{x}\phi_{B}\right]+\kappa_{BA}\partial_{x}^{2}\phi_{A}.
\end{align}
\end{subequations}
Numerical integration of Eq.~\eqref{eq:ptPhi2S} yields again stationary demixed states that transition to traveling ones with increasing nonreciprocity, as shown  in Fig. \ref{fig:2SupeCric}.

Following the procedure described above, we can obtain the one-mode approximation for Eq. \eqref{eq:ptPhi2S} as
\begin{subequations}
\label{eq:dtPhiFA2S}
\begin{align}
  \frac{d\hat{\phi}_{A}^{1}(t)}{dt}=&-\left(\alpha_{A}+|\hat{\phi}_{A}^{1}|^{2} \right)\hat{\phi}_{A}^{1} -(\kappa-\delta)\hat{\phi}_{B}^{1}, \\
  \frac{d\hat{\phi}_{B}^{1}(t)}{dt}=& -\left(\alpha_{B}+\yzh{|\hat{\phi}_{B}^{1}|^{2}}\right)\hat{\phi}_{B}^{1}-(\kappa+\delta)\hat{\phi}_{A}^{1},
\end{align}
\end{subequations}
where $\alpha_{\mu}=\chi_{\mu}+\gamma_{\mu}+(\phi_{\mu}^{0})^{2}$. In the amplitude-phase space, Eq. \ref{eq:dtPhiFA2S} can be rewritten as
\begin{subequations}
  \label{eq:dtPhiAP2S}
\begin{align}
\label{eq:dtPhiAA2S}
\frac{d \rho_{A}}{d t}=&-(\alpha_{A} +\rho_{A}^{2})\rho_{A}-(\kappa-\delta)\rho_{B}\cos\theta, \\
\label{eq:dtPhiAB2S}
  \frac{d \rho_{B}}{d t}=&-(\alpha_{B}+\rho_{B}^{2})\rho_{B}-(\kappa+\delta)\rho_{A}\cos\theta, \\
\label{eq:dtTheta2S}
  \frac{d \theta}{d t}=&\left[(\kappa-\delta)\rho_{A}^{-1}\rho_{B}+ (\kappa+\delta)\rho_{A}\rho_{B}^{-1} \right]\sin\theta, \\
  \label{eq:dtPhi2S}
  \frac{d \Phi}{d t}=&\left[(\kappa-\delta)\rho_{A}^{-1}\rho_{B}- (\kappa+\delta)\rho_{A}\rho_{B}^{-1} \right]\sin\theta\;.
\end{align}
\end{subequations}
While a detailed analysis of fixed points and their stability is more challenging in this case and is left for future work, the structure of the amplitude and phase equations is clearly very similar to the one discussed in the main text. In particular, a traveling modulation will again correspond to a solutions with $\sin\theta\not=0$ and
\begin{align}
&\left[(\kappa-\delta)\rho_{A}^{-1}\rho_{B}+ (\kappa+\delta)\rho_{A}\rho_{B}^{-1} \right]=0\;,\\
&\left[(\kappa-\delta)\rho_{A}^{-1}\rho_{B}- (\kappa+\delta)\rho_{A}\rho_{B}^{-1} \right]\not=0
\end{align}
to guarantee a finite value of \yzh{$v=\dot\Phi/2$}. These conditions immediately give $v=\pm\sqrt{\delta^2-\delta_c^2}$, which provides an excellent fit to the velocity obtained from simulations, as shown in Fig. \ref{fig:2SupeCric}c.

\section{Examples on nonreciprocity in physical systems}
The generic theory described in the main text suggests that nonreciprocity provides a generic mechanism for a
static-to-traveling transition in a field theory of coupled scalar fields that exhibit spatial domains or patterns.
In this section we
provide two examples of how this generic behavior can emerge from
specific microscopic models.

\subsection{Example 1 : microscopic nonreciprocity}
\begin{figure}[b]
\centering
\includegraphics[width=0.7\linewidth]{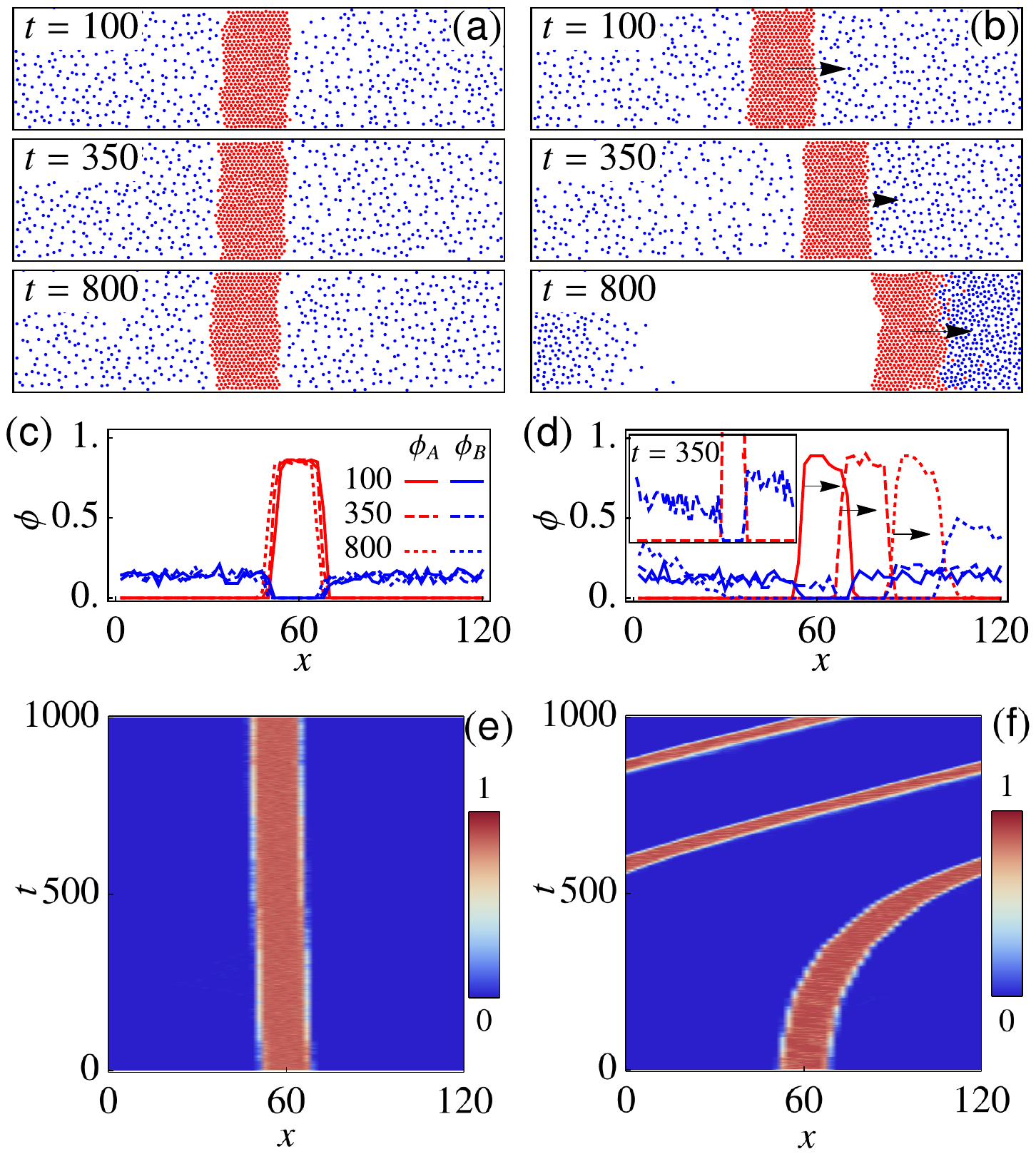}
\caption{\label{fig:AttrRepu} Transition from static (left panels: a, c, e) to traveling (right panels: b, d, f) states in a binary mixture of attractive-repulsive particles. We use $k_{AB}=-1$ for the left panels, and $k_{AB}=-2$ for the right panels. The red and blue particles represent species A and B, respectively. (a,b) Snapshots of particles at different time points. See also Movies S4-S5 for animation. (c,d) Distributions of particle densities in $x$ at the (c) static and (d) traveling states. The inset in (d) highlights the breaking of reflection symmetry of the blue line. The three different types of lines corresponds to the above three snapshots: solid--$t=100$, dashed--$t=350$, and dotted--$t=800$. We use red and blue lines to indicate densities of species $A$ and $B$, respectively. (e--f) \yzh{Spatiotemporal} patterns of $\phi_{A}(x)$ at the (e) static and (f) traveling states. In (c--f), each data point is obtained by measuring the density in a rectangle of width $\Delta x=2$, spanning the entire vertical direction, and centered at $x$, i.e. $\phi_{\mu}(x)=N_{\mu}(x)/(\Delta xL_{y})$, where $N_{\mu}$ is the number of $\mu$ particles whose $x$-positions are within $(x-\Delta x/2,x+\Delta x/2]$.}
\end{figure}

Let us consider a mixture of particles where the microscopic interactions between the two species are nonreciprocal. Instances of this include mixtures of chemically interacting colloids that are known to exhibit nonreciprocal effective interactions~\cite{Agudo-Canalejo2019, Saha2019} and chase-and-run dynamics in predator-prey systems~\cite{Mogilner2003, Chen2014}. Let us assume that such nonreciprocal interactions can be captured by pairwise additive short ranged central forces. Under such circumstances, the microdynamics of the particles' positions $\{\mathbf{r}_{i}^{\mu}\}_{i=1}^{N_{\mu}}$, with $\mu=A, B$, is governed by equations of the form
\begin{equation}
  \label{eq:ARMicro}
\partial _{t}\mathbf{r}_{i}^{\mu}=\sum_{j=1}^{N_{\mu}}\textbf{F}^{\mu\mu}(\textbf{r}_{i}^{\mu},\textbf{r}_{j}^{\mu})+\sum_{j=1}^{N_{\nu}}\textbf{F}^{\mu\nu}(\textbf{r}_{i}^{\mu},\textbf{r}_{j}^{\nu})+\bm{\eta
}_{i}^{\mu}.
\end{equation}
Here, $\bm{\eta}_{i}^{\mu}(t)$ is uncorrelated white noise with zero mean and variance $=\langle  \eta_{i\alpha}^{\mu}(t)\eta_{j\beta}^{\nu}(t')\rangle=\sqrt{2D}~\delta_{ij}\delta_{\mu\nu}\delta_{\alpha\beta}\delta(t-t')$, where
$\alpha,\beta$ denote Carthesian components.
$\textbf{F}^{\mu\mu}(\textbf{r}_{i}^{\mu},\textbf{r}_{j}^{\mu})$ and $\textbf{F}^{\mu\nu}(\textbf{r}_{i}^{\mu},\textbf{r}_{j}^{\nu})$ are the intra- and inter-species interactions of the form:
\begin{subequations}
\label{eq:F}
\begin{align}
  \textbf{F}^{\mu\mu}(\textbf{r}_{i}^{\mu},\textbf{r}_{j}^{\mu})&=k_{r}\Theta\left(\sigma_{r},r_{ij}^{\mu\mu}\right)\hat{\textbf{r}}_{ij}^{\mu\mu}+k_{\mu\mu}\Theta\left(\sigma_{\mu\mu},r_{ij}^{\mu\mu}\right)\hat{\textbf{r}}_{ij}^{\mu\mu}, \\
\textbf{F}^{\mu\nu}(\textbf{r}_{i}^{\mu},\textbf{r}_{j}^{\nu})&=k_{r}\Theta\left(\sigma_{r},r_{ij}^{\mu\nu}\right)\hat{\textbf{r}}_{ij}^{\mu\nu}+k_{\mu\nu}\Theta\left(\sigma_{\mu\nu},r_{ij}^{\mu\nu}\right)\hat{\textbf{r}}_{ij}^{\mu\nu},
  \end{align}
\end{subequations}
where $\mathbf{r}_{ij}^{\mu\nu}=\mathbf{r}_i^\mu-\mathbf{r}_j^\nu$ and $r_{ij}^{\mu\nu}=|\mathbf{r}_{ij}^{\mu\nu}|$. The first terms on the right-hand side of\eqref{eq:F} are  excluded-volume forces of strenght $k_r>0$ and range $\sigma_r$. The second terms represent longer-range repulsive or attractive  interactions of strength $k_{\mu\mu}$, and $k_{\mu\nu}$ and corresponding range $\sigma_{\mu\mu}$, $\sigma_{\mu\nu}$. All forces are piecewise linear, with $\Theta\left(\sigma,r_{ij}\right)=0$ for $r_{ij}>\sigma$ and $\Theta\left(\sigma,r_{ij}\right)=\sigma-r_{ij}$ when $r_{ij}<\sigma$. We assume that particles  $A$ are attracted to other $A$ particles ($k_{AA}<0$), so that species $A$ can self-aggregate and form a dense cluster. This qualitatively mimics the aggregation  driven by a negative $\chi_{A}$ in the model B discussed in the main text. To enforce nonreciprocity, we set $k_{AB}<0$ and $k_{BA}>0$, so species A (B) is attracted (repelled) by species B (A). As a proof of concept, we have simulated a mixture of $N_{A}=N_{B}=525$ particles in a periodic box of dimension $120\times 30$. The simulation parameters are: $\sigma_{r}=1$, $\sigma_{AB}=\sigma_{BA}=3$, $k_{r}=2000$, $D=1$, $k_{AA}=-2$, $k_{BA}=k_{BB}=1$, and vary $k_{AB}$ to study the pattern formation. Representative examples shown in Fig.~(\ref{fig:AttrRepu}) and \yzh{Movies S4-S5} indicate that this microscopic model does indeed exhibit the phenomenology discussed in the main body of the paper.

In order to sketch the connection between this microscopic model and the
generic theory considered in the main text, let us neglect the noise term in
Eq. (\ref{eq:ARMicro}) and consider the deterministic overdamped
microdynamics. The low density dynamics of the density fields
$\phi_\mu(\mathbf{r},t)=\sum _{i=1}^{N_\mu}\delta(\mathbf{r}-\mathbf{r}_i^\mu(t))$,
will generically be of the form
\begin{equation*}
\partial _{t}\phi _{\mu }\left( \mathbf{r},t\right) =-\bm\nabla _{r}\cdot \left[
\left\langle \mathbf{F}_{\mu }\left( \mathbf{r},t\right) \right\rangle \phi
_{\mu }\left( \mathbf{r},t\right) \right]
\end{equation*}%
where $\left\langle \mathbf{F}\right\rangle $ is the mean field force, which
can be estimated as%
\begin{eqnarray*}
\left\langle \mathbf{F}_{\mu }\right\rangle  &=&\int d\mathbf{r}^{\prime
}\left( k_{r}\Theta \left( \sigma _{r},\left\vert \mathbf{r}-\mathbf{r}%
^{\prime }\right\vert \right) +k_{\mu \mu }\Theta \left( \sigma _{\mu \mu
},\left\vert \mathbf{r}-\mathbf{r}^{\prime }\right\vert \right) \right)
\frac{\mathbf{r-r}^{\prime }}{\left\vert \mathbf{r}-\mathbf{r}^{\prime
}\right\vert }\phi _{\mu }\left( \mathbf{r}^{\prime },t\right)  \\
&&+\int d\mathbf{r}^{\prime }\left( k_{r}\Theta \left( \sigma
_{r},\left\vert \mathbf{r}-\mathbf{r}^{\prime }\right\vert \right) +k_{\mu
\nu }\Theta \left( \sigma _{\mu \nu },\left\vert \mathbf{r}-\mathbf{r}%
^{\prime }\right\vert \right) \right) \frac{\mathbf{r-r}^{\prime }}{%
\left\vert \mathbf{r}-\mathbf{r}^{\prime }\right\vert }\phi _{\nu }\left(
\mathbf{r}^{\prime },t\right)
\end{eqnarray*}%
Using the piecewise linear form of the forces, to lowest order in gradients
of the density, the mean field force can be evaluated to give%
\begin{equation*}
\left\langle \mathbf{F}\right\rangle =-R_{\mu \mu }\bm\nabla \phi _{\mu }\left(
\mathbf{r},t\right) -R_{\mu \nu }\bm\nabla \phi _{\nu }\left( \mathbf{r}%
,t\right)
\end{equation*}%
where
\begin{equation*}
R_{\mu \nu }=\left( k_{r}\sigma _{r}^{4}+k_{\mu \nu }\sigma _{\mu \nu
}^{4}\right) \frac{\pi }{6}
\end{equation*}%
Therefore, the equations of motion for the macrodynamics of this system are
\begin{equation*}
\partial _{t}\phi _{\mu }\left( \mathbf{r},t\right) =\bm\nabla \cdot \left[
R_{\mu \mu }\phi _{\mu }\left( \mathbf{r},t\right) \bm\nabla \phi _{\mu }\left(
\mathbf{r},t\right) \right] +\bm\nabla \cdot \left[ R_{\mu \nu }\phi _{\mu
}\left( \mathbf{r},t\right) \bm\nabla \phi _{\nu }\left( \mathbf{r},t\right) %
\right]
\end{equation*}
Focussing on the cross diffusion coefficient, note that in this model, the
analog of the reciprocal part of the cross diffusion coefficient  is $\kappa
_{\mu \nu }=k_{r}\sigma _{r}^{4}\frac{\pi }{6}\phi _{\mu }$, set by the
pairwise repulsive interactions and the analog of the nonreciprocal part is $%
\delta _{\mu \nu }=k_{\mu \nu }\sigma _{\mu \nu }^{4}\frac{\pi }{6}\phi _{\mu }$ and
scales with the microscopic nonreciprocal interaction
strength. Thus, this microscopic model yields a continuum theory that is
closely analogous to the generic theory we considered in the main text and, as shown in Fig. ~(\ref{fig:AttrRepu}),
exhibits the static to traveling pattern transition that is our central
result.

\subsection{Example 2 - Emergent nonreciprocity in mixtures of active
and passive particles}

As a second illustration of a microscopic model giving rise to
the phenomenology discussed in this work,  we consider a mixture
of active (A) and passive (P) Brownian particles \cite{Wysocki2016, Wittkowski2017}.

In the absence of interactions, the dynamics of each each Active Brownian Particle (ABP) is described by
\begin{equation}
\partial _{t}\mathbf{r}=v_{0}\mathbf{\hat{u}}  \label{AP1}
\end{equation}%
where $v_0$ is the propulsion speed, $\mathbf{\hat{u}}=\cos \theta \hat{x}+\sin \theta \hat{y}$ is the
direction of active motion which itself undergoes rotational diffusion,
i.e., $\partial_{t}\theta =\sqrt{2D_{R}}\xi \left( t\right) $ with $\xi
\left( t\right) $ is a delta-function correlated white noise of unit variance. The ballistic motion exhibited at short times by Eq. (\ref{AP1}) becomes diffusive at long times due to the rotational diffusion of the propulsion direction $\mathbf{\hat{u}}$ with a characteristic diffusion coefficient $v_{0}^{2}/2D_{R}$. In the presense of short range repulsive interactions given by some pairwise additive central potential $U_{AA}$, a collection of ABPs undergoes a well-studied athermal liquid-gas like
phase separation into a dense phase and a low density phase that has been dubbed MIPS (motility induced phase separation) \cite{Cates2015}. While the inherently nonequilibrium nature of this phase transition manifests itself in different interfacial phenomena~\cite{Wittkowski2014, Cates2019}, the bulk phenomenology nevertheless is approximately characterized by a supercritical Model B~\cite{Stenhammar2013, Redner2016}.

The second component of our mixture consists of passive Brownian
particles that exhibit diffusive dynamics and interact with each
other through a short range repulsive potential $U_{PP}$. This component by
itself will form a Brownian gas and its dynamics can be reasonably modeled
by a subcritical Model B dynamics \cite{Kardar2007, Doi2013}. We additionally assume that the two species are coupled through short range repulsive interactions given by a central potential of the form $U_{AP}$. Precisely such a model was studied through Brownian dynamics simulation in \cite{Wysocki2016} and it was shown that the active and passive particles phase separate and in certain regions of parameter space, the interface between the two species spontaneously starts to move. We demonstrate below that this phenomenon falls within the generic paradigm described in this work.

\begin{figure}[t]
\centering
\includegraphics[width=0.7\linewidth]{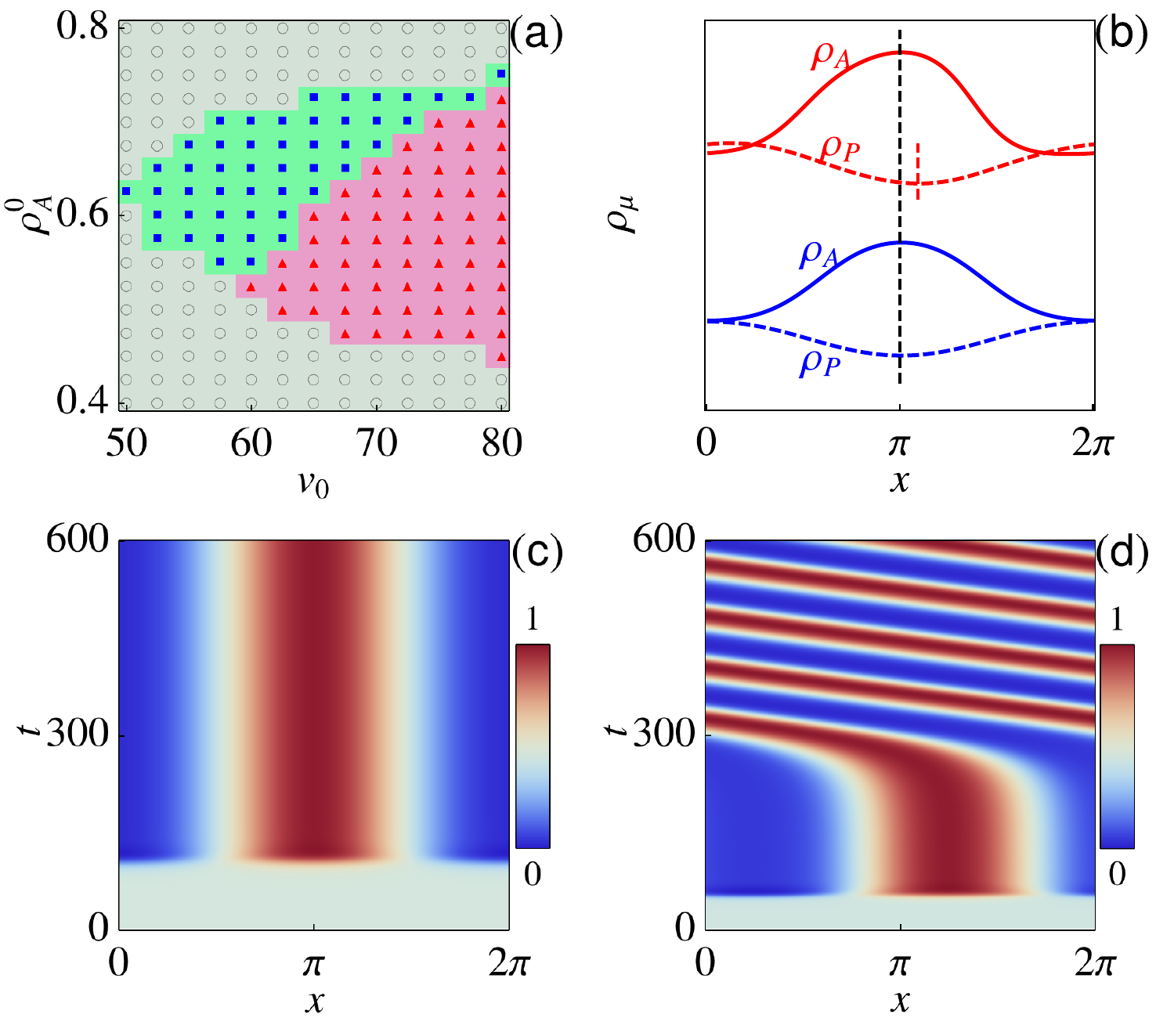}
\caption{\label{fig:ActiPass} (a) State diagram of the active-passive mixture spanned by $v_{0}$ and $\rho_{A}^{0}$ obtained from  numerical simulations of Eq.~\eqref{eq:ptRhoAP}. As in the model presented in the main text, the AP mixture exhibits  three states: a homogeneous state where active and passive particles are mixed (gray, circles), static state where active passive particles are concentrated in different spatial regions, hence demixed  (cyan, rectangles), and a state of demixed  traveling domains  (pink, triangles). (b) Examples of spatial variations of $\rho_{A}(x)$ (solid lines) and $\rho_{P}(x)$ (dashed lines) in the static (blue) and traveling states (red). (c--d) Spatiotemporal patterns of $\rho_{A}(x,t)$ in the (c) static and (d) traveling states. In (b--d), we have used $\rho_{A}^{0}=0.6$ and (c) $v_{0}=60$ and (d) $v_{0}=70$.}
\end{figure}

We stress that at the microscopic level, all interactions in the AP mixture are reciprocal. The corresponding coarse-grained equations as derived in \cite{Wittkowski2017} are given by
\begin{subequations}
\label{eq:ptRhoAP}
\begin{align}
\frac{\partial \rho _{A}}{\partial t}=& \nabla \cdot \lbrack D_{AA}\nabla
\rho _{A}+D_{AP}\nabla \rho _{P}]-\kappa \nabla ^{4}\rho _{A}\;, \\
\frac{\partial \rho _{P}}{\partial t}=& \nabla \cdot \lbrack \yzh{D_{PA}}\nabla
\rho _{A}+D_{PP}\nabla \rho _{P}]-\kappa \nabla ^{4}\rho _{P}\;,
\end{align}%
\end{subequations}
with  diffusion coefficients
\begin{subequations}
\label{AP2}
\begin{align}
D_{AA}=& a_{1}^{AA}\rho _{A}+\frac{v}{2D_{R}}\left(
v-a_{0}^{AA}v_{0}\rho _{A}\right)  \\
D_{AP}=& a_{1}^{AP}\rho _{A}-\frac{v}{2D_{R}}a_{0}^{AP}v_{0}\rho
_{A} \\
D_{PA}=& a_{1}^{PA}\rho _{P}, \\
D_{PP}=& a_{1}^{PP}\rho _{P}\;,
\end{align}
\end{subequations}%
Here, $v=v_{0}(1-a_{0}^{AA}\rho _{A}-a_{0}^{AP}\rho _{P})$ is the effective motility of active particles and $a_{i}^{\mu
\nu }$ coefficients depend on the pair potentials $U_{\mu \nu }$
and the statistics of interparticle collisions. While there are
subtle difference in the form of the direct diffusion coefficients of this
model as compared to our generic Model B, note that the direct diffusion coefficient of species A can be written in the form $D_{AA}\sim \frac{v^{2}\left( \rho_{A},\yzh{\rho _{P}}\right) }{2D_{R}}$, where $v\left( \rho _{A},\rho _{P}\right)=v_{0}(1-a_{0}^{AA}\rho _{A}-a_{0}^{AP}\rho _{P})$ is the effective density dependent motility of the active particles that are slowed down by collisions with both themselves and with the passive particles. Therefore MIPS driven here by the change in sign of $D_{AA}$ corresponds to the Hopf-bifurcation discussed in our generic theory and can be controlled by the density of either species.  Importantly, the cross diffusion coefficients calculated by \cite{Wittkowski2017} are indeed nonreciprocal. This nonreciprocity is emergent, in that it arises due to the statistics of the collisions rather than from nonreciprocity of interparticle interactions~\cite{Wittkowski2017,You2020}. The strength of nonreciprocity is controlled by $v_{0}$. Even though changing this parameter influences both the direct and cross diffusivities, $v_{0}$ can be considered analogue to $\delta$ in the generic model considered in the main text. Further, the Hopf bifurcation that controls MIPS rendering the field $\rho _{A}$ supercritical is now controlled by both the mean densities $\rho _{A}^{0}$ and $\rho _{P}^{0}$. We have studied numerically Eqs. (\ref{eq:ptRhoAP}) using $D_{R}=3$, $a_{0}^{AA}=1$, $a_{0}^{AP}=0.7$, $a_{1}^{AA}=a_{1}^{AP}=a_{1}^{PA}=a_{1}^{PP}=25$, and $\kappa=1000$. We have fixed $\rho _{P}^{0}=0.3$ and tuned $\rho _{A}^{0}$ and $v_{0}$, which serve respectively as the analog of the control parameter $\chi _{A}$ and $\delta$ in the coupled Model B considered in the main text. The phase behavior of this system is shown in Fig. \ref{fig:ActiPass} which reproduces the phenomenology discussed in the main body of the paper. Specifically, by increasing the degree of nonreciprocity $v_{0}$, one can see a transition from static out-of-phase pattern (blue rectangles in Fig. \ref{fig:ActiPass}a, blue lines in Fig. \ref{fig:ActiPass}b, Fig. \ref{fig:ActiPass}c) to a steady traveling pattern with broken parity (red triangles in Fig. \ref{fig:ActiPass}a, red lines in Fig. \ref{fig:ActiPass}b, Fig. \ref{fig:ActiPass}d).

\bibliography{SINonreciprocity}
\bibliographystyle{apsrev4-1}